\documentclass[aps,prd,10pt,notitlepage,nofootinbib,superscriptaddress]{revtex4-1}
\usepackage[utf8]{inputenc}
\usepackage{amsmath}
\usepackage{amssymb}
\usepackage{amsfonts}
\usepackage{newtxtext,newtxmath}
\usepackage{bm}
\usepackage{graphicx}

\usepackage[usenames,dvipsnames]{xcolor}
\usepackage{color}
\usepackage[colorlinks=true,linkcolor=blue,urlcolor=blue,citecolor=blue]{hyperref}

\usepackage{slashed}
\usepackage[english]{babel}
\usepackage{dcolumn}
\usepackage{blindtext}
\usepackage{epsfig}
\usepackage{pifont}
\usepackage{dsfont,mathrsfs}
\usepackage{cancel}
\usepackage{bigints}
\usepackage{accents}
\usepackage{soul}
\usepackage{multirow}
\usepackage{simpler-wick}
\usepackage{makecell}

\newcommand{\be}{\begin{equation}}
\newcommand{\ee}{\end{equation}}
\newcommand{\bea}{\begin{eqnarray}}
\newcommand{\eea}{\end{eqnarray}}
\newcommand{\ba}[1]{\begin{array}{#1}}
\newcommand{\ea}{\end{array}}
\newcommand{\nn}{\nonumber}

\newcommand{\om}{\omega}  
\newcommand{\FB}[1]{\left(#1\right)}

\newcommand{\vk}{\vec k}

\newcommand{\ensembleaverage}[1]{\left\langle#1\right\rangle}

\newcommand{\Ensembleaverage}[1]{\langle#1\rangle}

\newcommand{\SB}[1]{\left\{#1\right\}}
\newcommand{\TB}[1]{\left[#1\right]}

\newcommand{\mcT}{\mathcal{T}}
\newcommand{\mcTc}{\mathcal{T}_C}
\newcommand{\mcN}{\mathcal{N}}

\newcommand{\scrD}{\mathscr{D}}

\newcommand{\munu}{{\mu\nu}}

\newcommand{\alphabeta}{{\alpha\beta}}
\newcommand{\mnab}{{\mu\nu\alpha\beta}}
\newcommand{\IM}{\text{Im}}

\newcommand{\Tr}{\text{Tr}}

\newcommand{\ppll}{p_\parallel}

\newcommand{\kper}{k_\perp}
\newcommand{\pper}{p_\perp}

\newcommand{\identity}{\mathds{1}}

\newcommand{\psibar}{\overline{\psi}}

\newcommand{\wkl}{\omega_{kl}}

\def\be {\begin{equation}}
\def\ee {\end{equation}}
\def\nn {\nonumber}
\def\bea {\begin{eqnarray}}
\def\eea {\end{eqnarray}}

\begin{document}

\title{Quantum version of transport coefficients in Nambu--Jona-Lasinio model at finite temperature and strong magnetic field}

\author{Aritra Bandyopadhyay}
\affiliation{Institut für Theoretische Physik, Universität Heidelberg, Philosophenweg 16, 69120 Heidelberg, Germany}
\author{Snigdha Ghosh}
\affiliation{Government General Degree College Kharagpur-II, Paschim Medinipur - 721149, West Bengal, India}
 \author{Ricardo L.S. Farias}
 \affiliation{Departamento de Física, Universidade Federal de Santa Maria, Santa Maria, RS 97105-900, Brazil }
 \author{Sabyasachi Ghosh}
 \affiliation{ Indian Institute of Technology Bhilai, GEC Campus, Sejbahar, Raipur 492015, Chhattisgarh, India}

\date{Received: date / Revised version: date}

\begin{abstract}
We have estimated parallel and perpendicular components of electrical conductivity and shear viscosity of quark matter at finite magnetic field and temperature by using their one-loop Kubo expressions in the framework of Nambu--Jona-Lasinio (NJL) model.  At finite magnetic field, a non-trivial medium dependence of those quantities can be found. Previously these NJL-profiles have been addressed in relaxation time approximation, where cyclotron motion of quarks with medium dependent mass plays the key role. With respect to the earlier estimations, the present work provides further enriched profiles via Kubo framework, where field theoretical descriptions of quark transport with medium dependent mass and (Landau) quantized energy have been identified as the key ingredients. Hence the present study can be considered as the complete quantum field theoretical description of the transport coefficients in the framework of NJL model at finite temperature and magnetic field. 
\end{abstract}

\maketitle

\section{Introduction}

Production of strong magnetic fields in the early stages of relativistic heavy-ion collisions (HIC) is a longstanding topic, that is being extensively studied~\cite{Rafelski:1975rf,Kharzeev:2007jp}. The strengths of these produced fields have been estimated to be even larger than the strong-interaction scale $\Lambda^2_{\rm QCD} \simeq 0.06$~GeV$^2$ (e.g. Pb-Pb collisions at the Large Hadron Collider estimates $e B \sim 15 m^2_\pi \gg \Lambda^2_{\rm QCD}$; $m_\pi$ is the pion mass $\sim 0.135$ GeV)~\cite{Skokov:2009qp} which subsequently indicates that these fields heavily influence various observables in 
the hot and dense quark matter such as the quark condensates. Many such modifications have already been studied but the interpretations of those modifications in the system are still ambiguous~\cite{Tuchin:2013ie,Kharzeev:2015znc,Wang:2016mkm,Zhao:2019hta}. One of the biggest challenges to understanding these magnetic field-induced modifications is to grasp the time dependence of the produced magnetic field in the early stages of HIC. There are several schools of thought on this topic. Some of the preliminary studies indicated that the fields weaken fast as the system expands~\cite{Skokov:2009qp,Voronyuk:2011jd}. Then there are also studies that suggested that the induced electric currents in the expanding matter due to the produced magnetic fields can in turn produce magnetic fields again, overall changing the longevity of the early-produced fields~\cite{Tuchin:2013apa,McLerran:2013hla,Gursoy:2014aka,Tuchin:2015oka}. Very recent studies have again given emphasise on the short lifetime of the induced magnetic field~\cite{Wang:2021oqq} and further suggested that the search for any magnetic effects in the HIC would be highly challenging. The absence of CME signals in the isobar experiments~\cite{STAR:2021mii} also concurs with this inference. In Ref.~\cite{Shovkovy:2022bnd}, the validity of Ohm's law has been argued in view of the rapidly evolving quark matter produced in HIC and the behavior of the time dependent conductivity has been discussed. All these studies give strong indications that electrical conductivity can be a really important quantity in this scenario.
Hence, understanding the microscopic calculation of transport coefficients like electrical conductivity at finite magnetic fields might be considered an important topic to study. Electrical conductivity and other transport properties such as shear and bulk viscosity could have been computed unambiguously using Lattice QCD, a nonperturbative first-principles numerical method formulated in Euclidean space, if not for the crude inversion techniques to reconstruct the Minkowski spectral functions from Euclidean correlation functions. However, there are existing lattice QCD results for transport coefficients with SU(2) quenched simulations~\cite{Buividovich:2010tn} and with full QCD SU(2+1) simulations~\cite{Astrakhantsev:2019zkr}. There are also several other recent analytical nonperturbative studies of electrical conductivity at finite
magnetic fields ~\cite{Nam:2012sg,Hattori:2016cnt,Hattori:2016lqx,Harutyunyan:2016rxm,Kerbikov:2014ofa,Feng:2017tsh,Fukushima:2017lvb,Li:2018ufq,Das:2019wjg,Das:2019ppb,Ghosh:2019ubc,Satapathy:2021cjp,Satapathy:2021wex}. On the other hand, simulations of the field dynamics invariably involve solving relativistic and dissipative magnetohydrodynamics equations, which require other transport coefficients like shear and bulk viscosity. The magnetic field dependence of the shear viscosity was computed recently in Refs.~\cite{Li:2017tgi,Nam:2013fpa,Alford:2014doa,Tawfik:2016ihn,Tuchin:2011jw,Ghosh:2018cxb,Mohanty:2018eja,Dey:2019axu,Dey:2019vkn,Dash:2020vxk} and of the bulk viscosity in Refs.~\cite{Hattori:2017qih,Huang:2009ue,Huang:2011dc,Agasian:2011st,Agasian:2013wta}. 

If we analyse the framework of earlier calculations, then we can find two classifications:
\begin{enumerate}
    \item Relaxation time approximation (RTA) based kinetic theory expressions of electrical conductivity and shear viscosity, used in Refs.~\cite{Harutyunyan:2016rxm,Feng:2017tsh,Das:2019wjg,Das:2019ppb,Ghosh:2019ubc,Dey:2019axu,Dey:2019vkn} and Refs.~\cite{Li:2017tgi,Alford:2014doa,Tawfik:2016ihn,Tuchin:2011jw,Ghosh:2018cxb,Mohanty:2018eja,Dey:2019axu,Dey:2019vkn,Dash:2020vxk}~,
    \item Kubo formalism of the same, used in Refs.~\cite{Nam:2012sg,Hattori:2016cnt,Hattori:2016lqx,Kerbikov:2014ofa,Fukushima:2017lvb,Buividovich:2010tn,Astrakhantsev:2019zkr,Ghosh:2020wqx,Satapathy:2021cjp,Satapathy:2021wex}.
\end{enumerate}
Former formalism connect one-body kinematics with equilibrium distribution function and its deviation to 
many-body mechanics, appeared as thermodynamics and transport coefficients of the system. Next, the deviation
is obtained in terms of relaxation time $\tau_c$ by using the RTA based Boltzmann transport equation. 
On the other hand, Kubo relation defines transport coefficients as transportation probability field operators 
like shear stress and electro-magnetic current between two points.
So, the diagrammatic quantum field theory calculation becomes the scope of this framework. It is quite fascinating that these two completely different methodologies converge into exactly same expressions of transport coefficients in
absence of magnetic field, which is well established by earlier Refs.~\cite{Jeon:1994if,Fernandez-Fraile:2009eug,Ghosh:2016yvt,Ghosh:2014yea}. 
On the other hand it develops curiosity again, when finite magnetic field extension of RTA expressions~\cite{Ghosh:2018cxb,Mohanty:2018eja,Dey:2019axu,Dey:2019vkn,Dash:2020vxk}
and Kubo expressions~\cite{Ghosh:2020wqx,Satapathy:2021cjp,Satapathy:2021wex} don't coincide. It is because of the quantum field theoretical aspect of the magnetic field, which was missing in the RTA picture, and properly incorporated in the Kubo expressions.
Refs~\cite{Satapathy:2021cjp,Ghosh:2020wqx} have gone through leading order estimations of conductivity and viscosity components, where propagators at finite temperature ($T$) and magnetic field ($B$) have been considered. A rich quantum field theoretical (QFT) structure is noticed in the parallel and perpendicular components of those transport coefficients. 
In the present work we intend to explore the effects of that field theoretical structure on the transport coefficients within a particular system described by an effective QCD model, carrying a nontrivial $T$ and $B$ dependence. We have highlighted on the QFT modification of transport coefficients by comparing with their corresponding RTA expressions. These comparisons were already done in Refs.~\cite{Ghosh:2020wqx,Satapathy:2021cjp,Satapathy:2021wex} for massless fermionic or bosonic system in general but the present work has extended it into a more specific and realistic system - quark matter in the light of effective QCD model framework.  
 
Among the several existing effective QCD models, 
the Nambu--Jona-Lasinio (NJL) model~\cite{Nambu:1961tp,Nambu:1961fr} is well adopted for QCD phenomenology at finite temperature~\cite{Vogl:1991qt,Klevansky:1992qe,Hatsuda:1994pi,Buballa:2003qv}.
At finite $T$ and $B$, quark condensate and constituent quark mass become $T$, $B$ dependent functions via magneto-thermodynamical phase-space, built from the Landau quantization technique of thermal field theory (TFT).  
In the present work, we have used the NJL model of Refs.~\cite{Farias:2014eca,Farias:2016gmy}, a model that compliments the novel lattice QCD results~\cite{Bali:2011qj,Bali:2012zg,Bruckmann:2013oba,Endrodi:2013cs}, which first showed that strong magnetic fields have dramatic effects on the QCD phase diagram. These lattice simulations have found that though the magnitude of the light quark condensates increase with the magnetic field for low temperatures, they start to decrease for temperatures close to $T_{\rm pc}\simeq$ $0.16$ GeV, the region associated with the chiral symmetry restoration. Former case is connected with magnetic catalysis (MC) and latter case with inverse magnetic catalysis (IMC). These phenomena are well reviewed in Refs.~\cite{Gatto:2012sp,Miransky:2015ava,Andersen:2014xxa,Bandyopadhyay:2020zte,Andersen:2021lnk,Ayala:2021nhx}. 

As first attempts, Refs~\cite{Ghosh:2018cxb,Ghosh:2019ubc} have provided estimations of shear viscosity~\cite{Dey:2019axu,Mohanty:2018eja} and electrical conductivity~\cite{Harutyunyan:2016rxm,Dey:2019axu} in presence of magnetic field by using RTA expressions within the NJL model. But the field theoretical structure rich Kubo expressions~\cite{Satapathy:2021cjp,Ghosh:2020wqx} of the transport coefficients have never been explored within the NJL model incorporating IMC. In the present work we have specifically explored this.

The paper is organized as follows. Sec.~\ref{sec:NJLB} has gone through the main results of NJL model calculation at finite magnetic field. Next in Secs.~\ref{sec:CM}, \ref{sec:Kubo}, RTA and Kubo expressions of transport coefficients at finite magnetic field are respectively addressed. Then, our Kubo or field theory based NJL estimation of transport coefficients are plotted and their additional $T$, $B$ profiles with respect to their corresponding RTA or classical based estimations are discussed in Sec.~\ref{sec:Res}. At the end, Sec.~\ref{sec:sum} has summarized the study with the new findings.

\section{NJL model in presence of magnetic field}
\label{sec:NJLB}
For our purpose we have chosen here the isospin-symmetric two-flavor NJL model, 
whose Lagrangian density in presence of an {electromagnetic (EM)} field ($A^\mu$) is given by
\be
\mathcal{L}_\text{NJL}= -\frac{1}{4} F^{\mu\nu}F_{\mu\nu}
+ \bar{\psi}\left(\slashed{D}-\hat{m}\right)\psi
+ G\left[ (\bar{\psi}\psi)^2+(\bar{\psi}i\gamma_5{\vec\tau}\psi)^2\right],
\label{NJL_lag}
\ee
where $\psi$ depicts $u$ and $d$ quark fields iso-doublet, each being an $N_c (=3)-$plet, $N_c$ being the number of colors. { $\hat{m} = {\rm diag}(u,d)$ is the quark-mass matrix within exact isospin symmetry and hence in the rest of the paper we will work with the notation $m_u=m_d=m$}.
$D_\mu = i\partial_\mu - Q A_\mu$ is the covariant derivative, $A_\mu$ is the {(EM)} gauge field, $F_{\mu\nu} = \partial_\mu A_\nu - \partial_\nu A_\mu$, and $\vec \tau = (\tau^1, \tau^2, \tau^3)$ are the isospin Pauli matrices. $Q = {\rm diag}(q_f) ={\rm diag}(q_u = 2e/3, q_d =-e/3)$ is the charge matrix in the flavor space and $G$ is the coupling constant of the NJL model. Solution of this model in the mean-field approximation corresponds to the leading-order approximation in the $1/N_c$ expansion. { Since the NJL model is unrenormalizable,  in presence of an external magnetic field isolating the divergences from the vacuum structures comprised of various Landau levels requires extra care. Based on well  known results of~\cite{Ebert:1999ht,Ebert:2003yk,Menezes:2008qt}, and recently explored in Ref. \cite{Avancini:2019wed}, the magnetic field independent regularization (MFIR) present itself to be a satisfactory method to study NJL model in a constant and external magnetic field. The advantage of this regularization scheme is the complete separation of the magnetic field contributions from the vacuum or thermal contributions and subsequent regularization of the vacuum term using standard procedures. On the other hand, most of the non-MFIR based regularization schemes try to remove the divergences at each individual Landau levels by using sharp cutoff regulator functions depending on that particular Landau level. Sharply cutting off the divergences at each Landau levels generate abrupt transitions between Landau levels which can lead to unphysical results \cite{Allen:2015paa,Duarte:2015ppa}, as e.g. oscillations in the chiral quark condensate. Although use of smoother regulator functions improve the situation and also help us identify the unphysical oscillations from the possible physical ones \cite{Avancini:2019wed}.}

In the mean field approximation, the gap equation for the constituent quark mass $M$ 
at finite temperature $T$ and in the presence of a magnetic field~$B$ is given by
\bea
M = m - 2 G  \sum_{f=u,d}\langle \bar{\psi}_f\psi_f\rangle,
\label{Gap_B}
\eea
where $\langle \bar{\psi}_f\psi_f\rangle$ is the quark condensate of flavor~$f$. In presence of an external magnetic field $B$, $\langle \bar{\psi}_f\psi_f\rangle$ can be
written as a sum of three contributions~\cite{Ebert:2003yk,Menezes:2008qt,Farias:2014eca}: 
\bea
\langle \bar{\psi}_f\psi_f\rangle = \langle \bar{\psi}_f\psi_f\rangle^\text{vac} 
+ \langle \bar{\psi}_f\psi_f\rangle^B + \langle \bar{\psi}_f\psi_f\rangle^{T,B},
\label{gap}
\eea
with 
\begin{align}
\langle \bar{\psi}_f\psi_f\rangle^\text{vac} &= -\frac{MN_c}{2\pi^2}\left[\Lambda\sqrt{\Lambda^2+M^2} 
- M^2 \ln \left(\frac{\Lambda+\sqrt{\Lambda^2+M^2}}{M}\right) \right], 
\label{cond1} \\
\langle \bar{\psi}_f\psi_f\rangle^{B} &= -\frac{M|q_f|BN_c}{2\pi^2}\nn\\
&\left[\ln\Gamma(x_f) 
- \frac{1}{2}\ln(2\pi)+x_f-\frac{1}{2}\left(2x_f-1\right)\ln(x_f) \right], 
\label{cond2 } \\
\langle \bar{\psi}_f\psi_f\rangle^{T,B} &= \sum\limits_{l=0}^\infty \alpha_l \, %
\frac{M|q_f|BN_c}{2\pi^2}\int\limits_{-\infty}^\infty dp_z \,\frac{f_0(\om_{f,l})}{\om_{f,l}}~,
\label{cond3} 
\end{align}
where $\Gamma(x_f)$ is the Euler gamma function, $x_f = {M^2}/{(2|q_f|B)}$. Within MFIR regularization scheme, the vacuum term $\langle \bar{\psi}_f\psi_f\rangle^\text{vac}$ given in Eq.(\ref{cond1}) is ultraviolet divergent and we use a three-dimensional cutoff $\Lambda$. The parametrization of the model is determined treating the coupling $G$, the current quark mass $m$ and the cutoff $\Lambda$ as free parameters, which are fixed by fitting the vacuum values of the pion mass 
$m_{\pi}$, pion decay constant $f_{\pi}$ and quark condensate 
$\langle {\bar \psi}_f \psi_f \rangle$. In addition, in Eq.~(\ref{cond3}) $l$ represents Landau levels, with 
$\alpha_l = 2 - \delta_{l,0}$  being 
the spin degeneracy factor and $f_0(\om_{f,l})$ is the Fermi-Dirac distribution function:
\be
f_0 (\om_{f,l}) = \frac{1}{1 + e^{\beta \om_{f,l}}},
\label{FD-dist}
\ee
where 
\be
\om_{f,l} = (p_z^2+M^2+2l|q_f|B)^{1/2}.
\label{omega_f}
\ee
As mentioned earlier, in the present work, our aim is to include the effects of the inverse magnetic catalysis (IMC) phenomenon on the quasi particle effective/constituent quark mass $M$. Usual NJL model with a fixed coupling constant $G$, in the quasi-particle approximation is unable to describe IMC~\cite{Bali:2011qj,Bali:2012zg}. One effective way to resolve this issue is to impose that the coupling constant~$G$ of the model is $T-$ and $B-$dependent~\cite{Farias:2014eca,Farias:2016gmy,Tavares:2021fik}. Using $G(B,T)$ a precise description of the lattice results for the $u$ and $d$ quark condensates has been obtained within the NJL model with the parametrization~\cite{Farias:2016gmy}:
\bea
G(eB,T) = c(eB)\left[1-\frac{1}{1+e^{\beta(eB)[T_a(eB)-T]}} \right]+s(eB),
\label{GBT}
\eea
where $c(eB)$, $\beta(eB)$, $T_a(eB)$ and $s(eB)$ depend only on the magnitude of $B$ and their values for selected values of $B$ are given in Table~1\footnote{There is a typo in Table 1 of Ref.~\cite{Farias:2016gmy}, in the fifth column is $\beta \rightarrow \beta/10$.}
\begin{table}
\caption{Values of the fitting parameters in Eq.~\eqref{GBT}. Units are in appropriate 
powers of GeV.}

\label{glatnjl}
\begin{center}
\begin{tabular}{ccccc}
\hline\noalign{\smallskip}
$eB$ & $c$ &  $T_a$ & $s$   & $\beta$  \\ \hline\noalign{\smallskip}
0.0    &    0.900 &  0.168   &  3.731   &  400.00 \\\noalign{\smallskip}
0.2    &    1.226  &  0.168   &  3.262   &  340.12 \\\noalign{\smallskip}
0.4    &    1.769  &  0.169   &  2.294   &  229.88 \\\noalign{\smallskip}
0.6    &    0.741  &  0.156   &  2.864   &  144.01 \\\noalign{\smallskip}
0.8    &    1.289  &  0.158   &  1.804   &  115.06 \\\noalign{\smallskip}
\hline
\end{tabular}
\end{center}
\end{table}
The expression given in Eq.~\eqref{GBT} has been adopted in Ref.~\cite{Farias:2016gmy} for mere convenience since it is well adapted to fit LQCD results. All following numerical results refer to the parametrization adopted in~Ref.~\cite{Farias:2016gmy}. The described model with $G(B,T)$ was used to study the effects of a magnetic field on neutral pion mass in Refs.~\cite{{Avancini:2016fgq},Avancini:2018svs}, which subsequently produced results that agree with corresponding lattice QCD results~\cite{Brandt:2015hnz}. 
For further evidences and inputs about the IMC phenomenon readers can look into refs.~\cite{Bruckmann:2013oba,Ayala:2014gwa,Ayala:2014iba,Bandyopadhyay:2020zte,Bandyopadhyay:2020zte,Andersen:2021lnk,Ayala:2021nhx}.
 
\section{Relaxation time approximated (RTA) Expressions}
\label{sec:CM}
Our main aim is to apply the NJL model for calculating transport coefficients, based on quantum field theoretical calculation at finite temperature and magnetic field. To realize QFT contribution in transport coefficients for quark matter within NJL model, we will first revisit their expressions based on kinetic theory, as we intend to compare between them. First we will revisit the expressions of transport coefficients without Landau quantization, which we have broadly mentioned as RTA expressions throughout this paper.  Refs.~\cite{Ghosh:2018cxb,Ghosh:2019ubc} have already gone through the NJL model estimations of the RTA expressions of transport coefficients, whose framework will be quickly and briefly revisited in the next subsections for the sake of completeness. We want to emphasize here again that the present article is aimed to zoom into the quantum aspects of transportation and the transition from classical to quantum estimations, within the framework of NJL model.
%
%
\subsection{RTA expressions of Electrical Conductivity}
\label{sec:El_CM}
The conductivity tensor $\sigma^{ij}$ can be realized as a proportional connector between current density $J_i$ and electric field $E_j$ via macroscopic Ohm's law $J_i=\sigma^{ij}E_j$. In the microscopic picture of dissipation, we can assume that the equilibrium distribution function of quark at zero quark chemical potential, $f_0=\frac{1}{e^{\beta\om}+ 1}$ undergoes a small deviation $\delta f$. Therefore, one can express (dissipative) current density as~\cite{Dey:2019axu,Dey:2019vkn,lifshitz1995physical,Harutyunyan:2016rxm} 
\bea
J_i &=&4N_c\sum_{f=u,d}q_f\int \frac{d^3k}{(2\pi)^3}\frac{k_i}{\om}\delta f~,
\label{micro_E}
\eea
where $4N_c=2\times 2\times N_c$ are respectively spin, particle-anti-particle, and color degeneracy factors of medium constituent quark with electric charge $q_f$ and energy $\om=\{\vk^2+M^2\}^{1/2}$. Using relaxation time approximation (RTA) of the Boltzmann transport equation (BTE)
\be
{-q_f} ({\vec E} +\frac{\vec k}{\om}\times {\vec B})\cdot\nabla_k (f_0+\delta f) = \frac{-\delta f}{\tau_c}~,
\ee
one can know the form of $\delta f$~\cite{Dey:2019axu,Dey:2019vkn,lifshitz1995physical,Harutyunyan:2016rxm}. 

{ To do the calculations, electromagnetic field geometry should be fixed first. Here, we fix our external magnetic
field direction along the z-axis and electric field will be considered first along the x or y direction for calculating the perpendicular conductivity $\sigma_{xx}$ or $\sigma_{yy}$ and then subsequently along the z-axis for calculating the parallel conductivity $\sigma_{zz}$. So for the first case, $B$ is along z direction and $E$ is along x direction. Using this condition of electromagnetic field geometry in RTA based BTE,}
we will get~\cite{Dey:2019axu,Dey:2019vkn}
\bea
\delta f
={q_f}\tau_c\Big(\frac{k_x}{\om}+\frac{k_y}{\om}\frac{\tau_c}{\tau_{Bf}}\Big)E_x
\frac{1}{1+(\tau_c/\tau_{Bf})^2}~\beta f_0(1-f_0)~,
\label{df}
\eea
where $\tau_c$ is relaxation time and $\tau_{Bf}=\frac{\om}{{q_f}B}$ is inverse of cyclotron frequency. 
The second term in the RHS of Eq.~(\ref{df}) is related to the Hall conductivity and since the sign of $\tau_{Bf}$ for particles and anti-particles will be opposite, so the net value of the Hall component will be zero at zero quark chemical potential. At non-zero quark chemical potential, due to an imbalance of particle and anti-particle densities, one can get non-zero net Hall conductivity~\cite{Dash:2020vxk}. But that is not our matter of interest here as we restrict ourselves to the zero (net) quark density zone. Hence, from the first term of Eq.~(\ref{df}), we will get the perpendicular ($\sigma^{xx}$) component of conductivity tensor~\cite{Dey:2019axu,Dey:2019vkn,lifshitz1995physical,Harutyunyan:2016rxm}
\begin{align}
\sigma^{xx}_{\rm RTA}&=\sigma^{\perp}_{\rm RTA} = 4N_c\sum_{f=u,d}{q_f}^2 \beta \nn\\ &\int\frac{d^3k}{(2\pi)^3}
\tau_c\frac{1}{1+(\tau_c/\tau_{Bf})^2}\frac{k^2}{3\om^2} f_0(1\pm f_0)~.
\label{sxx_CM}
\end{align}
{ Similar to $\sigma^{xx}_{\rm RTA}$, one can get $\sigma^{yy}_{\rm RTA}$ by considering electric field along y direction instead of x direction and it can be easily found that both have same expressions. Both components can be called perpendicular components in general. 

Next for the parallel component of conductivity tensor ($\sigma^{zz}$), electric and magnetic field both are applied along the z-axis, so the Lorentz force will not work along the z direction. Hence $\sigma^{zz}$ comes out to be}
\be
\sigma^{zz}_{\rm RTA}=\sigma^{\parallel}_{\rm RTA} = 4N_c\sum_{f=u,d}{q_f}^2 \beta\int \frac{d^3k}{(2\pi)^3}
\tau_c\frac{k^2}{3\om^2} f_0(1\pm f_0)~.
\label{szz_CM}
\ee
This parallel component remains the same as we get isotropic expressions in the absence of the magnetic field. We will see later that this magnetic field independent behavior of parallel conductivity in classical case will be modified in the quantum picture.  


\subsection{Introducing Landau quantization in electrical conductivity: QM expressions}
\label{sec:El_QM}

After getting the RTA expressions of conductivity, here, we will add their quantum aspects. We know that Schrodinger's equation in presence of a magnetic field converts into a quantum harmonic oscillator-type problem with quantized energy, which is known as Landau quantization. For relativistic spin $0$ and spin $1/2$ particles, one has to respectively solve Klein-Gordon and Dirac's equations in a magnetic field to find out the quantized energy with Landau levels $l$. Here, we will straightaway impose the Landau quantization directly. For the rest of this manuscript, we have mentioned the Landau quantized RTA expressions as the QM (quantum mechanical) expressions. The main changes will be within the phase space integration $2\int \frac{d^3k}{(2\pi)^3}$, which will be changed to $\Big(\frac{|{q_f}|B}{2\pi} \Big)  \sum_{l=0}^\infty\alpha_l \int\limits^{+\infty}_{-\infty} \frac{dk_z}{2\pi}$ with spin degeneracy $\alpha_l=2-\delta_{l0}$ and within the energy $\om=\sqrt{\vk^2+M^2}$, which will now be quantized to $\om_{f,l}=\sqrt{\vk_z^2+M^2+2l{q_f}B}$. The perpendicular momenta of quark is considered as $k_\perp^2\approx (\frac{k_x^2+k_y^2}{2})=\frac{2l{q_f}B}{2} = lq_fB$. Based on these modifications, Eqs.~\eqref{sxx_CM} and \eqref{szz_CM} will now be expressed as :   
	\begin{align}
	\sigma^{xx}_{\rm QM}&=\sigma^{\perp}_{\rm QM} =2N_c \sum_{f=u,d}{q_f}^2 \beta\Big(\frac{|{q_f}|B}{2\pi} \Big) \sum_{l=0}^\infty \alpha_l 
	\nn\\ &\int\limits^{+\infty}_{-\infty} \frac{dk_z}{2\pi} \frac{{l|{q_f}|B}}{\om^2_{f,l}} \tau_c 
	\frac{1}{1+(\tau_c/\tau_{Bf})^2}
	f_0(\om_{f,l})[1-f_0(\om_{f,l})],
	\label{sxx_QM}
	\\
	\sigma^{zz}_{\rm QM}&=\sigma^{\parallel}_{\rm QM} = 2N_c \sum_{f=u,d}{q_f}^2 \beta\Big(\frac{|{q_f}|B}{2\pi} \Big) \sum_{l=0}^\infty \alpha_l 
	\nn\\ &\int\limits^{+\infty}_{-\infty} \frac{dk_z}{2\pi} \frac{k^2_z}{\om^2_{f,l}} \tau_c 
	f_0(\om_{f,l})[1-f_0(\om_{f,l})]~.
	\label{szz_QM}
	\end{align}
	

\subsection{RTA expressions of Shear viscosity}
\label{sec:Sh_CM}

After electrical conductivity, let us also quickly revisit the semi-classical expression of shear viscosity, whose detailed calculations can be found in Refs.~\cite{Dey:2019axu,Dey:2019vkn,lifshitz1995physical,Tuchin:2011jw,Ghosh:2018cxb,Mohanty:2018eja}.
The main motivation of this approach is to build a connection between macroscopic fluid description and microscopic kinetic theory description. Similar to electrical conductivity, which is realized as a proportionality constant between current density and electric field, shear viscosity also becomes the proportionality constant between viscous stress tensor and velocity gradient. 
{ In absence of magnetic field, the general form of Newton-Stoke relation is given by $\pi^{ij}=\eta {\cal U}^{ij}$ between viscous pressure $\pi^{ij}$ and velocity gradient ${\cal U}^{ij}= \left(\left(\frac{\partial u_{i}}{\partial x_{j}} 
+ \frac{\partial u_{j}}{\partial x_{i}}\right) + (2/3) \delta^{ij} \vec{\nabla} \cdot \vec{u} \right)$ with fluid velocity $u_i$.}

Now in presence of a magnetic field, one can construct five independent velocity gradient tensors, and hence, we get five shear viscosity components - $\eta_{0,1,2,3,4}$.
{ So, the magnetized medium modified Newton-Stoke relation can be written as
\be
\pi^{ij}=\sum_{n=0,..,4}\eta_n {\cal U}^{ij}_n~,
\ee
where velocity gradient ${\cal U}^{ij}_n$ will be constructed by unit vector of magnetic field along with the fluid velocity.}  
There are two different possible
sets of five independent trace less tensors, prescribed in Ref.~\cite{lifshitz1995physical}
and Refs.~\cite{Huang:2011dc,Huang:2009ue}. { Readers can find the detailed structure of ${\cal U}^{ij}_n$ in those Refs.~\cite{lifshitz1995physical,Huang:2011dc,Huang:2009ue}. In the microscopic picture of viscous dissipation, we can assume that due to velocity gradients ${\cal U}^{ij}_n$, the equilibrium distribution function $f_0$ undergoes a small deviation $\delta f_n$ along the directions of different velocity gradients ${\cal U}^{ij}_n$. Therefore, one can express the (dissipative) viscous pressure as~\cite{Dey:2019axu,Dey:2019vkn,lifshitz1995physical,Harutyunyan:2016rxm} 
\bea
\pi^{ij} &=&\sum_{n=0,..,4}4N_c\sum_{f=u,d}q_f\int \frac{d^3k}{(2\pi)^3}\frac{k^ik^j}{\om}\delta f_n~,
\label{micro_E}
\eea
Using RTA based BTE, one can know the form of $\delta f_n$~\cite{Dey:2019axu,Dey:2019vkn,Tuchin:2011jw,Ghosh:2018cxb,Mohanty:2018eja,Denicol:2018rbw,Dey:2019axu,Chen:2019usj} and get the
final microscopic expressions of $\eta_n$'s.} 
If we analyze the earlier existing references, we can find that Refs.~\cite{Dey:2019axu,Dey:2019vkn,Tuchin:2011jw,Ghosh:2018cxb,Mohanty:2018eja}
have adopted former set of tensors and Ref.~\cite{Denicol:2018rbw,Dey:2019axu,Chen:2019usj} have adopted latter set of tensors. Now five shear viscosity components, obtained from two different tensors are inter-connected~\cite{Dey:2019axu,Huang:2011dc,Huang:2009ue}, so one can proceed to use any one of the sets tensors. Here, we will use the tensors, prescribed by Refs.~\cite{Huang:2011dc,Huang:2009ue} and RTA expressions of $\eta_{0,1,2,3,4}$, whose detailed derivation can be found in Ref.~\cite{Dey:2019axu}.
Among them, $\eta_{3,4}$ is the Hall viscosity, which will be zero for vanishing quark chemical potential, just like the Hall conductivity. In the present work, we will focus on parallel ($\eta_{\parallel}$) and perpendicular ($\eta_{\perp}$) components of shear viscosity (considering the magnetic field is along the z-axis), which are related with $\eta_{0,2}$ as~\cite{Dey:2019axu,Critelli:2014kra,Finazzo:2016mhm}
\begin{align}
\eta^{\parallel}_{\rm RTA} &=\eta_{xzxz}=\eta_{yzyz}=\eta_0+\eta_2 = 4N_c \sum_{f=u,d}\frac{\beta}{15}
\nn\\
&\int \frac{d^3\vk}{(2\pi)^3}\left(\frac{\vk^2}{\om}\right)^2\tau_c
\frac{1}{\left\{1+(\tau_c/\tau_{Bf})^2\right\}}
\left\{f_0(1- f_0) \right\}~,\\
\eta^{\perp}_{\rm RTA}&=\eta_{xyxy}=\eta_0 = 4N_c \sum_{f=u,d}\frac{\beta}{15}
\nn\\
&\int \frac{d^3\vk}{(2\pi)^3}\left(\frac{\vk^2}{\om}\right)^2\tau_c
\frac{1}{\left\{1+4(\tau_c/\tau_{Bf})^2\right\}}
\left\{f_0(1- f_0) \right\}~,
\label{eta_pp}
\end{align} 
where RTA expression of $\eta_{2}$ is~\cite{Dey:2019axu,Dey:2019vkn,lifshitz1995physical,Tuchin:2011jw,Ghosh:2018cxb,Mohanty:2018eja}:
\begin{align}
\eta_2&=4N_c \sum_{f=u,d}\frac{3\beta}{15}\int \frac{d^3\vk}{(2\pi)^3}\left(\frac{\vk^2}{\om}\right)^2\tau_c \nn\\
&\frac{(\tau_c/\tau_{Bf})^2}{\{1+4(\tau_c/\tau_{Bf})^2\}\{1+(\tau_c/\tau_{Bf})^2\}} 
\left\{f_0(1- f_0) \right\}~.
\label{eta2_CM}
\end{align}

\section{Kubo Expressions}
\label{sec:Kubo}

In Sec.~\ref{sec:CM}, we have discussed about RTA based expressions of transport coefficients like shear viscosity and electrical conductivity. In this section we will discuss an alternative methodology, where one can find a quantum field theoretical structure of these transport coefficients. Owing to the Kubo relation, one can represent different transport coefficients as zero momentum limit of thermal correlator for corresponding field operator. These correlators can be expressed as one-loop self energy diagrams by considering the kinetic/free part of Lagrangian density. With further inclusion of thermal width $\Gamma_c$ (which can also be considered as the inverse of relaxation time $\tau_c$ in RTA approach) in propagator, one can get a non-divergent expression of transport coefficients. In absence of magnetic field, this procedure yields one loop Kubo expressions of different transport coefficients, which are exactly same as their RTA expressions. But in presence of magnetic field the RTA and Kubo expressions differ from each other~\cite{Satapathy:2021cjp,Ghosh:2020wqx,Satapathy:2021wex}. In the present article, we have estimated those Kubo expressions within the NJL model. In the present section we discuss those expressions for electrical conductivity and shear viscosity. For more details on the following formalism one can look into Refs.~\cite{Satapathy:2021cjp,Ghosh:2020wqx,Satapathy:2021wex}.
\subsection{Kubo expressions of Electrical conductivity}
\label{sec:El_QFT}

In the Kubo formalism, the electrical conductivity tensor $\sigma^{\mu\nu}$ is calculated from the long-wavelength limit of the in-medium {(EM)} spectral function $\rho^{\mu\nu}$ as
\begin{eqnarray}
	\sigma^{\mu\nu} = \lim\limits_{\vec{q}=\vec{0}, q^0\to0} \rho^{\mu\nu}(q).
	\label{eq.sigma}
\end{eqnarray}
Here, the spectral function $\rho^{\mu\nu}$ is related to the the Fourier transform of the vector current-current correlator given by
\begin{eqnarray}
	\rho^{\mu\nu}(q) &=& \tanh\FB{\frac{q^0}{2T}}\IM ~i \int\! d^4x e^{i q \cdot x} \ensembleaverage{ \mathcal{T}_cJ^{\mu}(x)J^{\nu}(0)}_{11},
	\label{spec1}
\end{eqnarray}
in which $\Ensembleaverage{...}$ denotes the ensemble average, $\mathcal{T}_c$ is the time ordering with respect to symmetric Schwinger-Keyldish contour $C$ in the complex time plane as used in the real time formalism (RTF) of finite temperature field theory, and, $11$ refers to the fact that the two points (in time ordering) are on the real horizontal segment of the contour $C$. Now, using vector current $J^\mu = -\psibar \gamma^\mu Q \psi$ of 2-flavor NJL quark matter for charge $Q$ and Wick's theorem in Eq.~\eqref{spec1}, we get
\begin{align}
\rho^{\mu\nu}(q)&= \tanh\FB{\frac{q^0}{2T}}\IM ~i \times \nn\\
&\int\! d^4x e^{i q \cdot x} \wick[offset=1.2em]{ \ensembleaverage{\mcTc \c2\psibar(x)Q\gamma^\mu \c1 \psi(x) \c1 \psibar(y)Q\gamma^\nu \c2 \psi(y)}_{11}} \\
	&= \tanh\FB{\frac{q^0}{2T}}\IM ~i \times\nn\\
	&\int\! d^4x e^{i q \cdot x}(-)\Tr_\text{d,c,f}\SB{\gamma^{\mu}S_{11}(x,y)Q\gamma^{\nu}S_{11}(y,x)Q}~,	\label{eq.z.1}
\end{align}
where $S_{11}(x,y) = \wick[offset=1.2em]{\ensembleaverage{\mcTc \c\psi(x) \c\psibar(y) }_{11}}$ is the 11-component of the thermo-magnetic real time quark propagator in coordinate space and $\Tr_\text{d,c,f}\{...\}$ refers to the trace taken over Dirac, color and flavor space. The propagator $S_{11}(x,y)$ is diagonal in both the flavor and color space as $S_{11}(x,y) = diag\FB{S_{11}^u(x,y),S_{11}^d(x,y)}\otimes\identity_\text{color}$ and the diagonal flavor components read
\begin{eqnarray}
	S^f_{11}(x,y) = \Phi^f(x,y)\int\!\!\!\frac{d^4p}{(2\pi)^4}e^{-ip\cdot(x-y)}\FB{-iS^f_{11}(p;m)}
	\label{eq.sxy}
\end{eqnarray}
where, $\Phi^f(x,y); f\in\{u,d\}$ is the gauge dependent phase factor and $S^f_{11}(p)$ is the 11-component of the momentum space thermo-magnetic quark propagator in RTF of flavor $f$, explicitly given by~\cite{Ayala:2003pv,Schwinger:1951nm}
\begin{eqnarray}
	S^f_{11}(p;m) &=& \sum_{l=0}^{\infty}(-1)^l\exp\FB{\frac{p_{\perp}^2}{|q_fB|}} \mathscr{D}_{lf}(p)\nn\\
	&&\TB{\frac{-1}{\ppll^2-M_{lf}^2+i\epsilon} - \xi(p^0)2\pi i\delta(\ppll^2-M_{lf}^2)}
	\label{eq.sp}
\end{eqnarray}
in which $\xi(x)=\Theta(x)f_0(x)+\Theta(-x)f_0(-x)$, $f_0(x)$ denotes the Fermi-Dirac thermal distribution function already defined in Eq.~\eqref{FD-dist}, $l$ is the Landau level index, $M_{lf}=\sqrt{M^2+2lq_fB}$ and $\scrD_l(p)$ is 
\begin{align}
	&\scrD_{lf}(p) = \FB{\cancel{p}_\parallel+M}\Bigg[\FB{\mathds{1}+sgn(q_f)i\gamma^1\gamma^2}L_l\FB{-\frac{2p_{\perp}^2}{|q_fB|}} \nn\\
		&- \FB{\mathds{1}-sgn(q_f)i\gamma^1\gamma^2}L_{l-1}\FB{-\frac{2p_{\perp}^2}{|q_fB|}}\Bigg] - 4\cancel{p}_\perp L^1_{l-1}\FB{-\frac{2p_{\perp}^2}{|q_fB|}} 
	\label{eq.Dl}
\end{align}
with the convention $L_{-1}(z) = L_{-1}^1(z) = 0$ for the Laguerre polynomials $L_l$. While writing the propagator in presence of magnetic field in $\hat{z}$ direction, we took $p_{\parallel,\perp}^\mu = g_{\parallel,\perp}^\munu p_\nu$ with $g_\parallel^\munu = diag(1,0,0,-1)$ and $g_\perp^\munu = diag(0,-1,-1,0)$ so that $\ppll^2=(p_0^2-p_z^2)$ and $\pper^2 = -(p_x^2+p_y^2) < 0$. Substituting Eq.~\eqref{eq.sp} into Eq.~\eqref{eq.sxy} followed by substituting in Eq.~\eqref{spec1}, we get~\cite{Satapathy:2021cjp,Ghosh:2020wqx,Satapathy:2021wex}
\begin{align}
	&\rho^{\mu\nu}( q_0, \vec{q}=\vec{0}) 
	= \lim\limits_{\Gamma\to0} \tanh\FB{\frac{q^0}{2T}}\sum_{f\in\{u,d\}} \sum_{l=0}^{\infty}\sum_{n=0}^{\infty}\int\!\!\!\frac{d^3k}{(2\pi)^3} \nn\\
	&\frac{1}{4\omega_{kl}\omega_{kn}}
	\big\{-f_0(\omega_{kl}) - f_0(\omega_{kn}) + 2f_0(\omega_{kl})f_0(\omega_{kn}) \big\} \nn \\
	&\times \Big[\mathcal{N}^{\mu\nu}_{lnf}(k_0=-\omega_{kl})\frac{\Gamma}{\Gamma^2+(q_0 - \omega_{kl} + \omega_{kn})^2}   \nn\\
	&+\mathcal{N}^{\mu\nu}_{lnf}(k_0 = \omega_{kl})\frac{\Gamma}{\Gamma^2 + (q_0+\omega_{kl}-\omega_{kn})^2}\Big],
	\label{eq.rhoB0}
\end{align}
where,
\begin{align}
	\mathcal{N}^{\mu\nu}_{lnf}(k) &= -N_cq_f^2(-1)^{l+n}\exp\FB{\frac{2k_{\perp}^2}{|q_fB|}}	\nn\\
	&\Tr_\text{d} \SB{\gamma^\mu \scrD_n(k) \gamma^\nu \scrD_l(k)}.
\end{align}

Now substituting Eq.~\eqref{eq.rhoB0} into \eqref{eq.sigma}, we obtain the conductivity tensor as 
\begin{align}
	\sigma^{\mu\nu} &= \frac{\partial\rho^{\mu\nu}}{\partial q_0}\Big|_{\vec{q}\to \vec{0},q_0 = 0} 
	= \lim\limits_{\Gamma\to 0} \sum_{f\in\{u,d\}} \sum_{l=0}^{\infty}\sum_{n=0}^{\infty}\frac{1}{2T}\nn\\
	&\int\!\!\!\frac{d^3k}{(2\pi)^3}\frac{1}{4\omega_{kl}\omega_{kn}}
	\frac{\Gamma}{\Gamma^2 + (\omega_{kl} -\omega_{kn})^2} \nn \\
	& \times~ \big\{-f_0(\omega_{kl}) - f_0(\omega_{kn}) + 2f_0(\omega_{kl})f_0(\omega_{kn})\big\} \nn\\
	&\Big[\mathcal{N}^{\mu\nu}_{lnf}(k,k)\big|_{k_0 = \omega_{kl}} + \mathcal{N}^{\mu\nu}_{lnf}(k,k)\big|_{k_0 = -\omega_{kl}}\Big].
	\label{rhoBTrans}
\end{align}
Having obtained the conductivity tensor, it is easy to extract the parallel and perpendicular components of conductivity using 
\begin{eqnarray}
	\sigma^{\parallel,\perp} = \mathcal{P}_{\mu\nu}^{\parallel,\perp}\sigma^{\mu\nu}
	\label{proj1}
\end{eqnarray}
where, the projectors $\mathcal{P}_{\mu\nu}^{\parallel,\perp}$ are given by 
\begin{eqnarray}
	\mathcal{P}_{\mu\nu}^\parallel &=& b^{\alpha}b^{\beta}\Delta_{\alpha\mu}\Delta_{\beta\nu},\\
	\mathcal{P}_{\mu\nu}^\perp &=& -\frac{1}{2}\Xi^{\alpha\beta}\Delta_{\alpha\mu}\Delta_{\beta\nu},
\end{eqnarray}
in which, $b^\mu = \frac{1}{2B}\varepsilon^{\mu\nu\alpha\beta}F_{\nu\alpha}u_{\beta}$, $u^\mu$ is the fluid four-velocity, $F_{\mu\nu} = \FB{\partial_{\mu}A_{\nu,\text{ext}} - \partial_{\nu}A_{\mu,\text{ext}}}$ is the field strength tensor, anti-symmetric in Lorentz indices, $b^{\mu\nu}= \varepsilon^{\mu\nu\alpha\beta}b_{\alpha}u_{\beta}$ and $\Xi^{\mu\nu} = \Delta^{\mu\nu} + b^{\mu}b^{\nu}$. In the local rest frame (LRF) of the fluid, $u^{\mu}_\text{LRF} = (1,0,0,0)$ and $b^{\mu}_\text{LRF} = (0,0,0,1)$ points along the direction of external magnetic field. Substituting Eq.~\eqref{rhoBTrans} into Eq.~\eqref{proj1} and performing the $d^2\kper$ integral analytically we finally arrive at
\begin{align}
	\sigma^{\parallel,\perp}_{\rm Kubo} &= \frac{N_c}{T}\sum_{f\in\{u,d\}} \sum_{n=0}^{\infty}\sum_{l=0}^{\infty}\int_{-\infty}^{+\infty}\frac{dk_z}{2\pi}\frac{1}{4\omega_{kl}\omega_{kn}} 
	\frac{\Gamma}{(\omega_{kl}-\omega_{kn})^2 + \Gamma^2} \nn\\
	&\big\{-f_0(\omega_{kl}) - f_0(\omega_{kn} )+ 2f_0(\omega_{kl})f_0(\omega_{kn})\big\}\mathcal{N}_{lnf}^{\parallel,\perp} (k_z)
	\label{kubo_ec}
\end{align}
where,
\begin{eqnarray}
	\mathcal{N}_{lnf}^\parallel(k_z) &=& q_f^2 \frac{|q_fB|}{\pi} \Big[ 4|q_fB| n\delta_{l-1}^{n-1} - \FB{\delta_{l}^{n}+\delta_{l-1}^{n-1}} 
	\FB{ k_z^2 -m^2 + \omega_{kl}^2}\Big],\\
	\mathcal{N}_{lnf}^\perp(k_z) &=& q_f^2\frac{|q_fB|}{\pi}\FB{\delta_{l}^{n-1} + \delta_{l-1}^{n}} \FB{k_z^2 + m^2 -\omega_{kl}^2}.
\end{eqnarray}
It is to be noted that $\delta_{-1}^{-1}=0$ which follows from the convention $L_{-1}(z)=L_{-1}^1(z) =0$ used in Eq.~\eqref{eq.Dl}.

{
Few comments on using the thermo-magnetic quark propagator of Eq.~\eqref{eq.sxy} and \eqref{eq.sp} while calculating the in-medium spectral function $\rho^\munu$ in Eq.~\eqref{spec1} are in order here. We note that, the local current $J^\mu(x)=-\psibar(x)\gamma^\mu Q\psi(x)$ appearing in Eq.~\eqref{spec1} consists of Heisenberg quark field $\psi(x)$ corresponding to the full interacting Hamiltonian of the model that includes the effects of effective `strong' interaction as well as of static (due to external magnetic field) and dynamical EM interaction at finite temperature and/or density. This in turn implies that, the quark propagator $S_{11}(x,y) = \wick[offset=1.2em]{\ensembleaverage{\mcTc \c\psi(x) \c\psibar(y) }_{11}}$ appearing in the expression of the spectral function $\rho^\munu$ in Eq.~\eqref{eq.z.1} should be the complete `dressed' propagator of the quarks carrying the effects of all the interactions. Now, the quark propagator in Eq.~\eqref{eq.sp} which is used to calculate the spectral function, does contain both the effects of effective `strong' as well as static EM interaction due to external magnetic field. In particular, the temperature and magnetic field dependent constituent quark mass $M=M(T,eB)$ entering in the expression of the propagator, captures the effect of `strong' interaction; whereas, the use of Schwinger proper-time formalism along with RTF takes into account the effects of static EM interaction due to external magnetic field and finite temperature to all orders. However, in this work, we have neglected the effect of dynamical EM interaction at finite temperature in the quark propagator which is of the order  $\alpha_\text{em}T^2$~\cite{Bellac:2011kqa} where $\alpha_\text{em} = 1/137$ is the fine structure constant. Clearly, the effect of dynamical EM interaction is sub-leading as compared to the effect of `strong' interaction as well as to the external magnetic field scale $eB \gg \alpha_\text{em}T^2$ for the temperature and magnetic field ranges considered in this work.  
}

From Eq.~(\ref{kubo_ec}) one can clearly see the difference between Kubo and RTA expressions. In Sec.~\ref{sec:El_QM}, we have explicitly imposed Landau quantization in RTA expressions (Eqs.~(\ref{sxx_QM}), (\ref{szz_QM})) but in the Kubo expression, Landau quantization is implicitly accounted for, originating from the thermo-magnetic quark propagator.

\subsection{Kubo expressions for Shear viscosity}

Similar to the previous subsection, the viscosity components can analogously calculated from the Kubo formalism. The fourth rank viscosity tensor $\mathcal{V}^{\mu\nu\alpha\beta}$ is calculated from the long-wavelength limit of the in-medium spectral function $\rho^{\mu\nu\alpha\beta}$ of the energy-momentum tensor (EMT) as
\begin{eqnarray}
	\mathcal{V}^{\mu\nu\alpha\beta} = \lim\limits_{\vec{q}=\vec{0}, q^0\to0} \rho^{\mu\nu\alpha\beta}(q).
	\label{eq.v}
\end{eqnarray}
Here, the spectral function $\rho^{\mu\nu\alpha\beta}$ is related to the the Fourier transform of the two-point EMT-EMT correlator given by
\begin{eqnarray}
	\rho^{\mu\nu\alpha\beta}(q) &=& \tanh\FB{\frac{q^0}{2T}}\IM ~i \int\! d^4x e^{i q \cdot x} \ensembleaverage{ \mathcal{T}_cT^{\mu\nu}(x)T^{\alpha\beta}(0)}_{11},
	\label{spec3}
\end{eqnarray}
Now, for 2-flavor NJL quark matter in MFA, the EMT is given by,
\begin{eqnarray}
T^\munu &=& \frac{1}{4}\FB{\psibar\gamma^\mu D^\nu\psi+D^{*\nu}\psibar\gamma^\mu\psi} - \frac{1}{2}g^\munu \bar{\psi}\left(\slashed{D}-M\right)\psi  + (\mu \leftrightarrow \nu). \label{eq.emt.dirac.B}
\end{eqnarray} 
Substituting Eq.~\eqref{eq.emt.dirac.B} into Eq.~\eqref{spec3}, we get~\cite{Satapathy:2021cjp,Ghosh:2020wqx,Satapathy:2021wex}
\begin{align}
	&\rho^{\mu\nu\alpha\beta}( q_0, \vec{q}=\vec{0}) 
	= \lim\limits_{\Gamma\to0} \tanh\FB{\frac{q^0}{2T}}\sum_{f\in\{u,d\}} \sum_{l=0}^{\infty}\sum_{n=0}^{\infty}\int\!\!\!\frac{d^3k}{(2\pi)^3}\nn\\
	&\frac{1}{4\omega_{kl}\omega_{kn}}
	\big\{-f_0(\omega_{kl}) - f_0(\omega_{kn}) + 2f_0(\omega_{kl})f_0(\omega_{kn}) \big\} \nn \\
	& \times \Big[\mathcal{N}^{\mu\nu\alpha\beta}_{lnf}(k_0=-\omega_{kl})\frac{\Gamma}{\Gamma^2+(q_0 - \omega_{kl} + \omega_{kn})^2}  \nn\\ &+\mathcal{N}^{\mu\nu\alpha\beta}_{lnf}(k_0 = \omega_{kl})\frac{\Gamma}{\Gamma^2 + (q_0+\omega_{kl}-\omega_{kn})^2}\Big],
	\label{eq.rhoB1}
\end{align}
where,
\begin{align}
	\mcN_{lnf}^\mnab(k) &= -\frac{1}{4} \Big[
	\mcT^{\mu\alpha}_{ln}(k)k^\nu k^\beta 
	- g^\munu \Big\{ \mcT^{\sigma\alpha}_{ln}(k)k_\sigma k^\beta 
	- m \mcT^{\alpha}_{ln}(k) k^\beta \Big\}\nn\\
	&- g^\alphabeta \Big\{ \mcT^{\mu\sigma}_{ln}(k)k_\sigma k^\nu 
	- m \mcT^{\mu}_{nl}(k) k^\nu \Big\} \nn \\
	& + g^\munu g^\alphabeta \Big\{ \mcT^{\sigma\rho}_{ln}(k)k_\sigma k_\rho 
	-2m \mcT^{\sigma}_{ln}(k)k_\sigma + m^2 \mcT_{ln}(k)\Big\}
	\Big] \nn\\
	&+ (\mu \leftrightarrow \nu) +  (\alpha \leftrightarrow \beta) 
	+ (\mu \leftrightarrow \nu, \alpha \leftrightarrow \beta),
\end{align}
in which, 
\begin{align}
	\mcT^\munu_{ln}(k) &= N_c(-1)^{l+n}\exp\FB{\frac{2k_{\perp}^2}{|q_fB|}}  \Tr \TB{ \gamma^\mu \scrD_n(k)\gamma^\nu \scrD_l(k) } \\
	\mcT^\mu_{ln}(k) &= N_c(-1)^{l+n}\exp\FB{\frac{2k_{\perp}^2}{|q_fB|}} \Tr \TB{ \scrD_n(k)\gamma^\mu \scrD_l(k) }, \\
	\mcT_{ln}(k) &= N_c(-1)^{l+n} \exp\FB{\frac{2k_{\perp}^2}{|q_fB|}} \Tr \TB{  \scrD_n(k) \scrD_l(k) }
\end{align}

Finally substituting Eq.~\eqref{eq.rhoB1} into \eqref{eq.v}, we obtain the viscosity tensor as 
\begin{align}
	\mathcal{V}^{\mu\nu\alpha\beta} &= \frac{\partial\rho^{\mu\nu}}{\partial q_0}\Big|_{\vec{q}\to \vec{0},q_0 = 0} = \lim\limits_{\Gamma\to 0} \sum_{f\in\{u,d\}} \sum_{l=0}^{\infty}\sum_{n=0}^{\infty}\frac{1}{2T}\nn\\
    &\int\!\!\!\frac{d^3k}{(2\pi)^3}\frac{1}{4\omega_{kl}\omega_{kn}}\frac{\Gamma}{\Gamma^2 + (\omega_{kl} -\omega_{kn})^2}\nn\\
	& \times~ \big\{-f_0(\omega_{kl}) - f_0(\omega_{kn}) + 2f_0(\omega_{kl})f_0(\omega_{kn})\big\} \nn\\
	&\Big[\mathcal{N}^{\mu\nu\alpha\beta}_{lnf}(k,k)\big|_{k_0 = \omega_{kl}} + \mathcal{N}^{\mu\nu\alpha\beta}_{lnf}(k,k)\big|_{k_0 = -\omega_{kl}}\Big].
	\nonumber\\
	\label{rhoBTrans.v}
\end{align}
Having obtained the viscosity tensor, it is easy to extract the parallel and perpendicular components of shear viscosity using 
\begin{eqnarray}
	\eta^{\parallel,\perp} = \mathcal{P}_{\mu\nu\alpha\beta}^{\parallel,\perp}\mathcal{V}^{\mu\nu\alpha\beta}
	\label{proj2}
\end{eqnarray}
where, the projectors $\mathcal{P}_{\mu\nu\alpha\beta}^{\parallel,\perp}$ are given by 
\begin{eqnarray}
	\mathcal{P}^\perp_\mnab &=& \frac{1}{4}
	\FB{\Xi^\sigma_\mu \Xi^\rho_\nu - \frac{1}{2}\Xi^{\sigma\rho}\Xi_\munu}
	\FB{\Xi_{\sigma\alpha} \Xi_{\rho_\beta} - \frac{1}{2}\Xi_{\sigma\rho}\Xi_\alphabeta} ~,\\
	\mathcal{P}^\parallel_\mnab &=& -\frac{1}{2} \Xi^\sigma_\mu b_\nu\Xi_{\sigma\alpha}b_\alpha ~.
\end{eqnarray}

Finally substituting Eq.~\eqref{rhoBTrans.v} into Eq.~\eqref{proj2} and performing the $d^2\kper$ integral analytically we finally arrive at
\begin{align}
	\eta^{\parallel,\perp}_{\rm Kubo} &= \frac{N_c}{T}\sum_{f\in\{u,d\}} \sum_{n=0}^{\infty}\sum_{l=0}^{\infty}\int_{-\infty}^{+\infty}\frac{dk_z}{2\pi}\frac{1}{4\omega_{kl}\omega_{kn}} \nn\\
	&\frac{\Gamma}{(\omega_{kl}-\omega_{kn})^2 + \Gamma^2}
	\big\{-f_0(\omega_{kl}) - f_0(\omega_{kn} ) \nn\\ 
	&+ 2f_0(\omega_{kl})f_0(\omega_{kn})\big\}\widetilde{\mathcal{N}}_{lnf}^{\parallel,\perp} (k_z),
	\label{eta_QFT}
\end{align}
where,
\begin{eqnarray}
	\tilde{\mcN}_{lnf}^\perp(k_z) &=& -2 \mathcal{D}^{(2)}_{ln} (\wkl^2-k_z^2-m^2) ~,\\
	\tilde{\mcN}_{lnf}^\parallel(k_z) &=&  8 \mathcal{B}^{(4)}_{ln}+\mathcal{C}^{(2)}_{ln} (\wkl^2+k_z^2-m^2)\nn\\
	&&+2 \mathcal{D}^{(0)}_{ln} k_z^2 (\wkl^2 - k_z^2 - m^2) + 4 \mathcal{E}^{(2)}_{ln} k_z^2,
	\label{N_Sh_B}
\end{eqnarray}
in which,
\begin{eqnarray}
	\mathcal{B}_{ln}^{(4)} &=& \frac{|q_fB|^3}{32\pi}nl\FB{2\delta_{l-1}^{n-1}+\delta_{l-1}^{n}+\delta_{l-1}^{n-2}}, \\
	\mathcal{C}_{ln}^{(2)} &=& -\frac{(q_fB)^2}{16\pi}\Big\{(2n+1)\delta_{l}^{n}+(n+1)\delta_{l}^{n+1} 
	+ n\delta_{l}^{n-1} \nn\\ 
	&& + (2n-1)\delta_{l-1}^{n-1}+n\delta_{l-1}^{n} 
	+ (n-1)\delta_{l-1}^{n-2} \Big\}, \\
	\mathcal{D}_{ln}^{(0)} &=& -\frac{|q_fB|}{8\pi}\FB{\delta_{l}^{n-1} + \delta_{l-1}^{n}}, \\
	\mathcal{D}_{ln}^{(2)} &=& \frac{(q_fB)^2}{16\pi}\Big\{(2n-1)\delta_{l}^{n-1}+n\delta_{l}^{n} 
	+ (n-1)\delta_{l}^{n-2} \nn\\
	&&+ (2n+1)\delta_{l-1}^{n}+(n+1)\delta_{l-1}^{n+1} 
	+ n\delta_{l-1}^{n-1} \Big\}, \\
	\mathcal{E}_{ln}^{(2)} &=& - \frac{(q_fB)^2}{16\pi}(l+n)\FB{\delta_{l-1}^{n-1}+\delta_{l-1}^{n}+\delta_{l}^{n-1}+\delta_{l}^{n}}.
\end{eqnarray}
At the end, let us put our working formulae in Table~(\ref{tab2}).
\begin{table}
\caption{Equation numbers of $\eta^{\perp,\parallel}$ and $\sigma^{\perp,\parallel}$, which will be the working
formulae in the result section.}
\label{tab2}
\begin{center}
\begin{tabular}{c|c|c}
\hline   
 & $\eta^{\perp,\parallel}$ &  $\sigma^{\perp,\parallel}$  \\ \hline  
~~~CM~~~    &    ~~~Eqs.~(\ref{eta_pp})~~~ &  ~~~Eqs.~(\ref{sxx_CM}), (\ref{szz_CM})~~~  \\\hline  
QM    &    - &  Eqs.~(\ref{sxx_QM}), (\ref{szz_QM})   \\\hline  
QFT    &    Eq.~(\ref{eta_QFT})  & Eq.~(\ref{kubo_ec})   \\  
\hline
\end{tabular}
\end{center}
\end{table}

\section{Results}
\label{sec:Res}

In this section we will discuss the numerical results of Kubo expressions within NJL model and compare them with previously explored RTA and QM expressions to reveal their additional contributions. Since NJL model at finite temperature and magnetic field provides us magneto-thermodynamic phase-space information of QCD from non-perturbative to perturbative domain, our plan is to identify them. For this purpose we have to take the massless quark results, which are close to pQCD results, as a reference to identify the non-perturbative effects by looking at the differences between massless and NJL model results. As a result, we will be careful in our graphical representations to reveal the different changes like classical to quantum, or pQCD to non-pQCD transformations. 

\begin{figure}[t]
   	\centering
   	\includegraphics[scale=0.5]{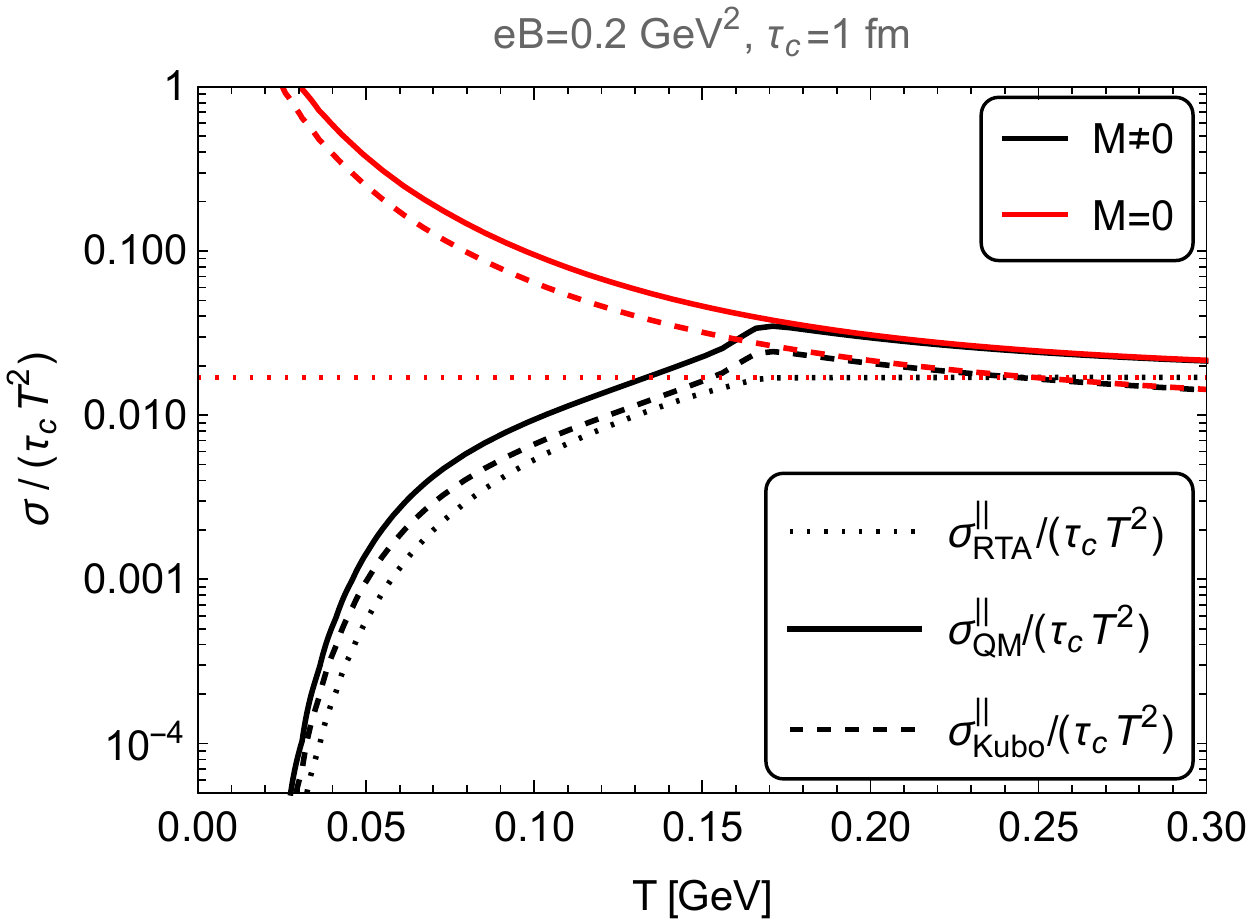} 
   	\includegraphics[scale=0.5]{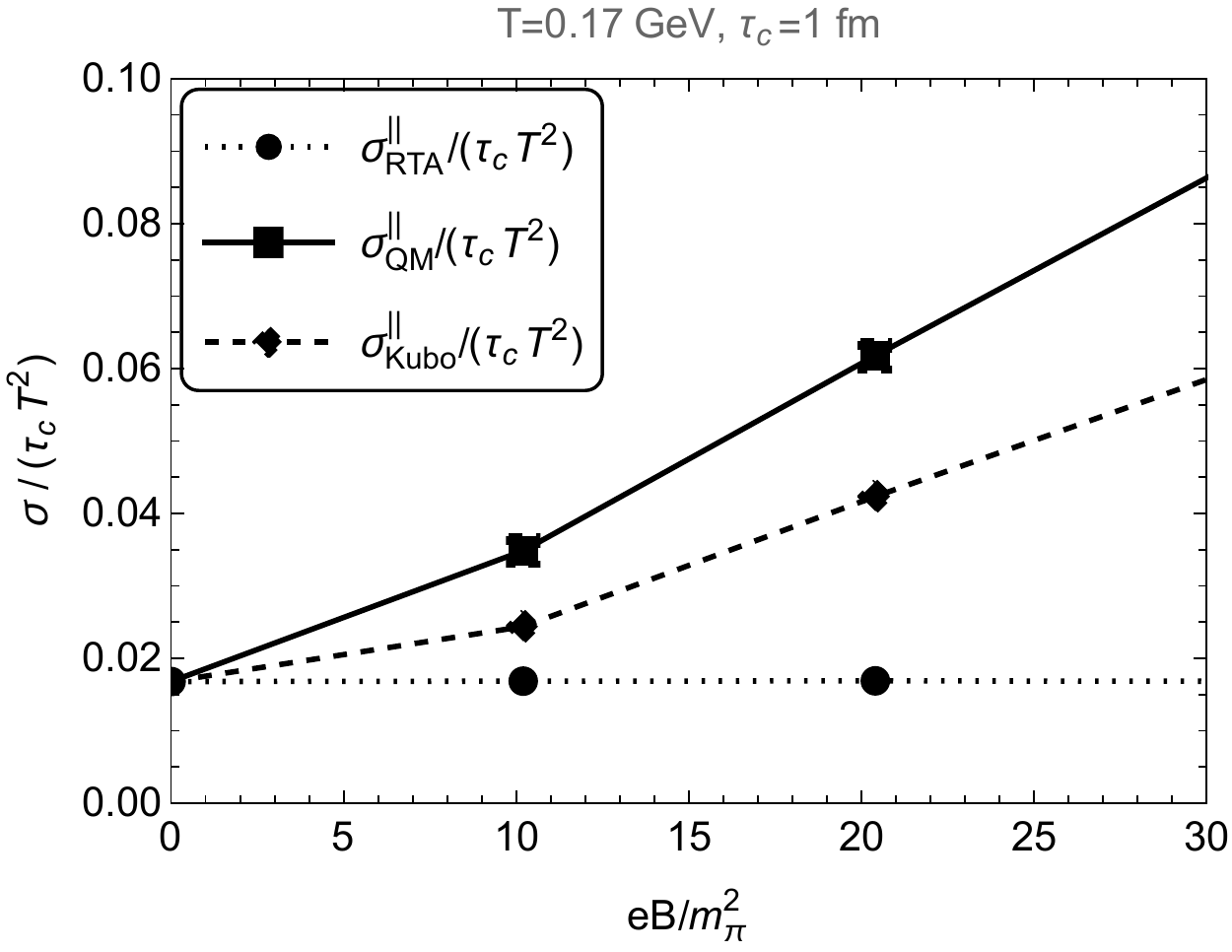}
   	\includegraphics[scale=0.5]{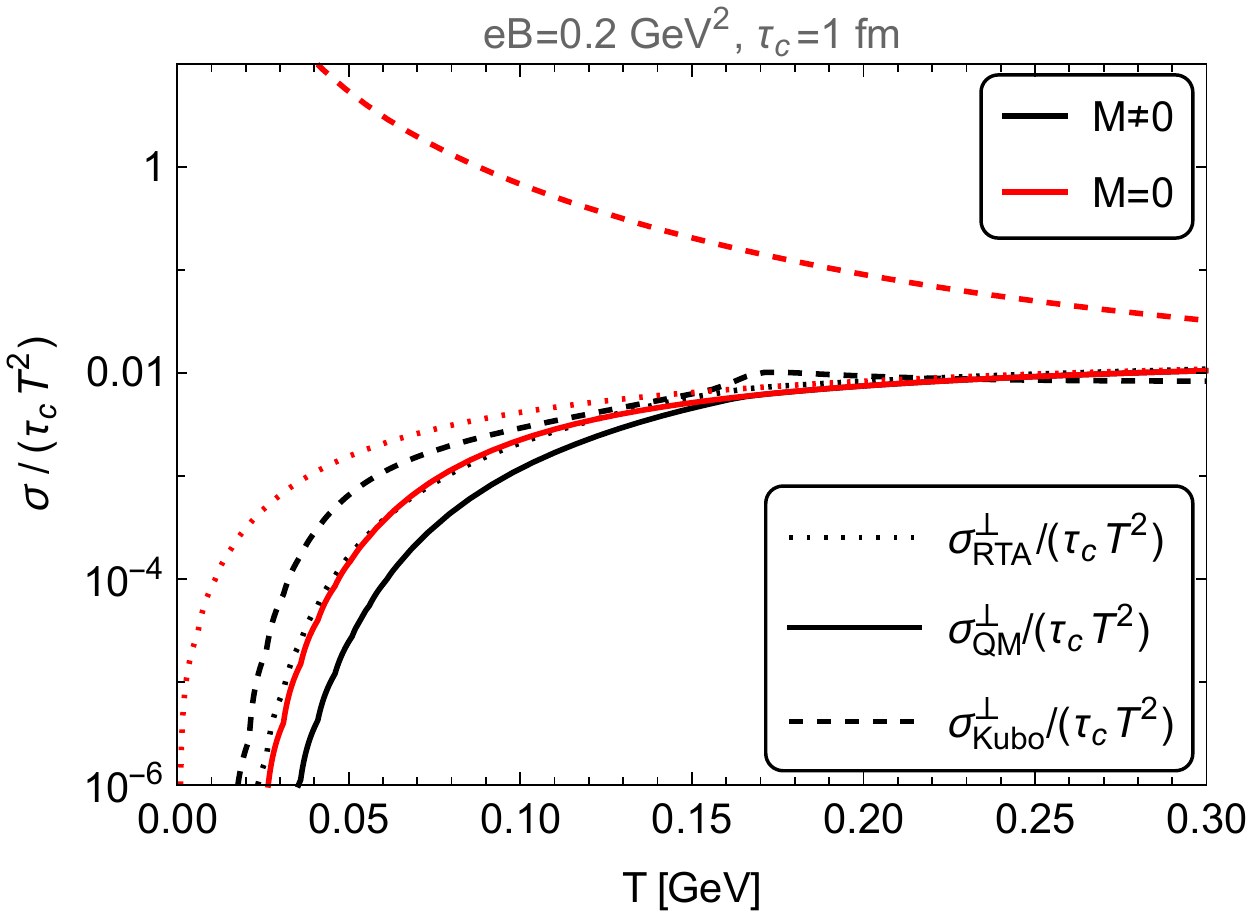} 
   	\includegraphics[scale=0.5]{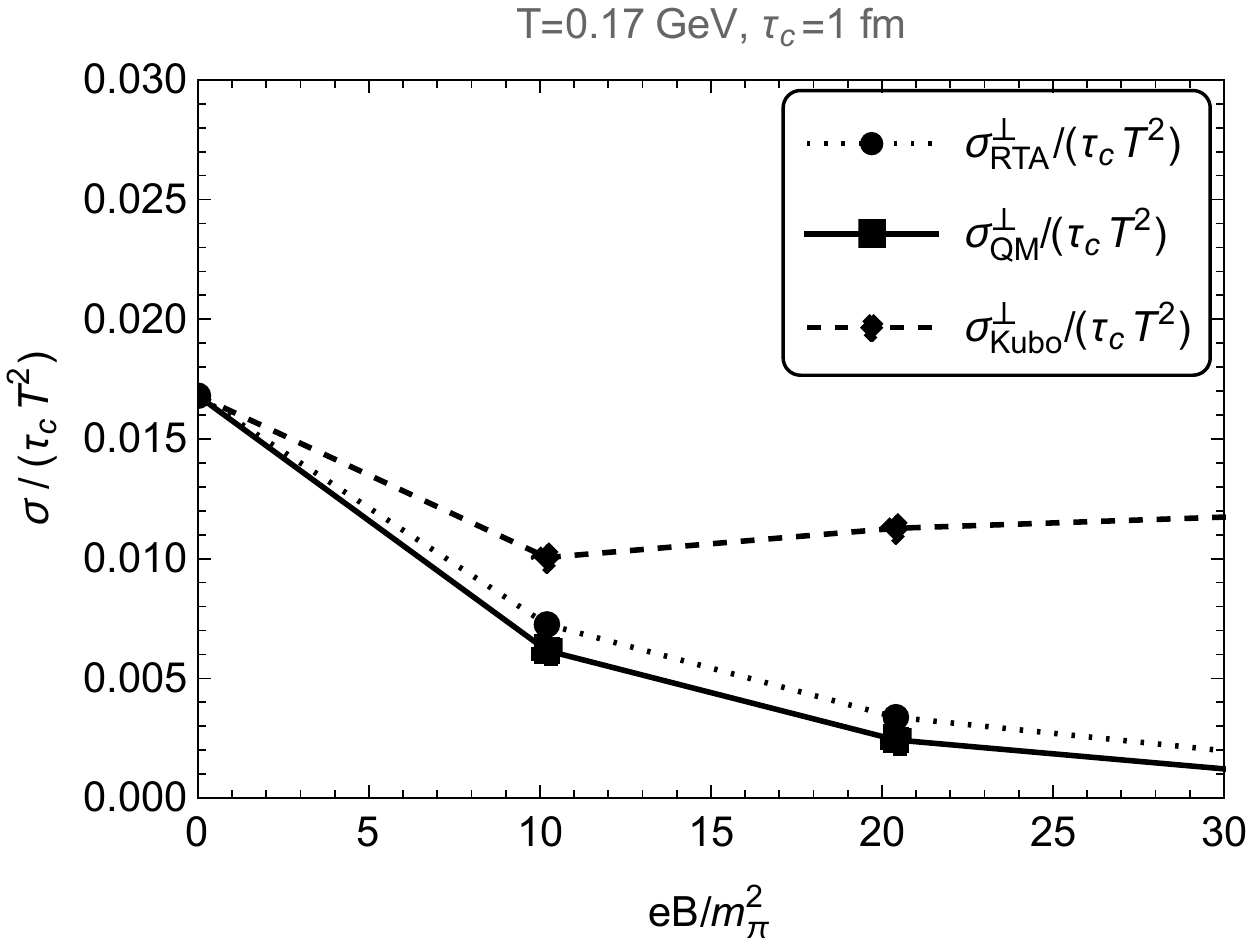}
   	\caption{Parallel and perpendicular conductivity of CM (dotted line), QM (solid line), QFT (dash line) curves vs temperature (left) and magnetic field (right).}
   	\label{fig:el_TB}
    \end{figure}

Let us start with electrical conductivity curves, shown in Fig.~\ref{fig:el_TB} where RTA, QM, and Kubo results are denoted by dotted, solid, and dashed curves. RTA curves are built from Eqs.~(\ref{sxx_CM}) and (\ref{szz_CM}) by using NJL based quark mass $M(T,B)$, obtained from Eq.~(\ref{Gap_B}). If we take the massless limit, parallel conductivity becomes proportional to $T^2$. So by choosing appropriate normalized quantities, we can get $T$ and $eB$ independent values~\cite{Dey:2019axu,Dey:2019vkn}, e.g. :
\bea 
\frac{\sigma^{\parallel}}{\tau_cT^2}&=&\frac{g}{18}\sum_{f=u,d} q_f^2=\frac{6}{18}\frac{5e^2}{9}\approx 0.017 ~.
\eea 
Horizontal dotted line in the left-upper panel of Fig.~\ref{fig:el_TB} indicate this $T$ and $eB$ independent nature of mass less RTA curves. Similar to Stephan-Boltzmann (SB) lines in the plots of lattice quantum chromodynamics (LQCD) based thermodynamics~\cite{Bali:2011qj,Bali:2012zg}, we may treat these horizontal curves as a reference line of transport coefficients. One can notice that when we consider $M(T,B)$, we get suppressed values of $\frac{\sigma^{\parallel}}{\tau_cT^2}$ concerning its massless values. The same trend is noticed in LQCD thermodynamics~\cite{Bali:2011qj,Bali:2012zg}. This suppression can be realized as a non-pQCD effect, observed in both thermodynamics and transport coefficients of QCD medium. Usually, it is the LQCD calculations~\cite{Bali:2011qj,Bali:2012zg}, effective QCD models like NJL model~\cite{Farias:2014eca,Farias:2016gmy} or LQCD mapping quasi-particle model~\cite{Dey:2019vkn}, which attempt to map non-pQCD contributions at finite temperature and magnetic field.  

Next, we go for the QM and Kubo curves of parallel components by using Eq.~(\ref{szz_QM}) and Eq.~(\ref{kubo_ec}), where Landau quantization is incorporated. We draw QM and Kubo curves separately to establish that Kubo curves cover all the field theoretical quantum effects, whereas QM curves cover a partial aspect of it. So introducing Landau quantization by hand in RTA expressions, through which we get the QM curves, might not be recommended for getting the full quantum effect in transport coefficients. Here, our final focal interest will be the Kubo curves. We present the QM curves as a bridge between the RTA and Kubo curves.  
At low $T$ in the left-upper panel of Fig.~\ref{fig:el_TB}, we notice that massless QM and Kubo curves deviate from their horizontal RTA curve. In Ref.~\cite{Bali:2011qj}, a similar pattern has observed for massless RTA and QM thermodynamical quantities like entropy density. Landau quantization plays an important role for this deviation from the horizontal line in both cases - the entropy density graph in Ref.~\cite{Bali:2011qj} and the plot of $\frac{\sigma^{\parallel}}{\tau_cT^2}$ in the left-upper panel of Fig.~\ref{fig:el_TB}. When we used $M(T, B)$ in the QM and Kubo expressions of $\sigma^{\parallel}$, we again get suppressed results in low $T$ concerning their massless results. Right-upper panel of Fig.~\ref{fig:el_TB} shows that the enhancement of quantum estimations of $\sigma^{\parallel}$ compared to its classical values will increase with the magnetic field. Classical and quantum curves will tend to merge at high $T$ and low $eB$ domains, which one can understand as the classical domain. On the other hand, quantum estimations of $\sigma^{\parallel}$ remain larger than classical estimations in the low $T$ and high $eB$ domain, which can be understood as the quantum domain. { This phenomena is quite interesting and could possibly connect to the magneto-resistance in the domain of condensed matter physics~\cite{Boris1,Boris2}, but a systematic and comparative study in future would be required to better comment on this topic.} 

Next, we explore the perpendicular component of the electrical conductivity. From Eq.~(\ref{sxx_CM}), one can identify the effective relaxation time~\cite{Dey:2019axu,Dey:2019vkn}
\be 
\tau^{\perp}_{cf}=\frac{\tau_c}{1+(\tau_c/\tau_{Bf})^2}~,
\ee 
with an approximate value inverse cyclotron frequency of massless quark matter~\cite{Dey:2019axu,Dey:2019vkn} :
\bea 
\tau_{Bf} &=& \frac{7\zeta(4)T}{2\zeta(3)q_fB}= \frac{3.15 T}{q_fB}~.
\label{tB_m0}
\eea 
Due to this term, $\sigma^{\perp}/(\tau_c T^2)$ curve of massless matter does not remain horizontal along $T$ and $eB$ axes like $\sigma^{\parallel}/(\tau_c T^2)$. Replacing the zero mass by NJL based $M(T,B)$, we will get further suppression due to non-pQCD effect. Red and black dotted lines for massless and $M(T,B)$ in left-lower panel of Fig.~\ref{fig:el_TB} display this fact. Next, QM (solid) and Kubo (dashed) curves of $\sigma^{\perp}/(\tau_c T^2)$ are generated by using Eqs.~(\ref{sxx_QM}) and (\ref{kubo_ec}) respectively. They are different in massless case~\cite{Ghosh:2020wqx} as well as for NJL based $M(T,B)$. The main reason is the transformation to an effective relaxation time expression:
\be 
\frac{\tau_c}{1+(\tau_c/\tau_{Bf})^2}\rightarrow \frac{\Gamma}{(\om_{f,l}-\om_{f,n})^2+\Gamma^2}~.
\ee 
In both the cases of RTA and QM, we consider classical concept of cyclotron motion using $\tau_B$, while in Kubo case, a transition between two energy levels, separated by unit Landau level difference~\cite{Ghosh:2020wqx} has come into the picture to describe the perpendicular conductivity components. Similar to parallel conductivity, perpendicular conductivity increases with respect to its classical values as we increase the magnetic field. One can observe this in the right-lower panel of Fig.~\ref{fig:el_TB}. Reader may notice that the right panel graphs, plotted against $B$-axis, carry few data points because of unavailability of LQCD data of $B$. However, it does not hinder guessing the approximated trend of the curves. We would also like to mention at this point that generating the QM or Kubo results at $eB \to 0$ is an impossible task (as we have to consider summing over an infinite number of Landau levels). But the coincidence of all three curves at $eB=0$ is expected and alternatively checked. 
\begin{figure}[t]
   	\centering
   	\includegraphics[scale=0.5]{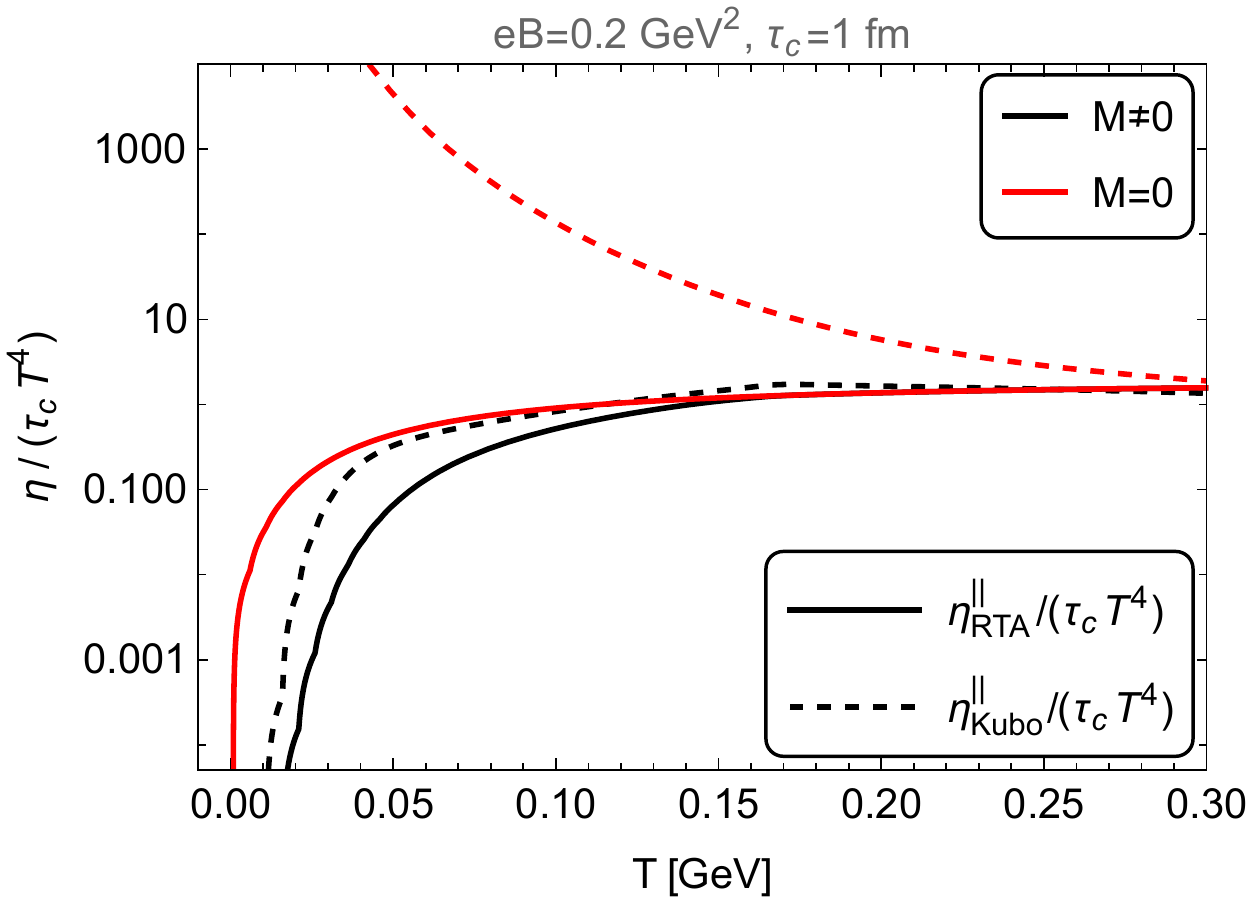} 
   	\includegraphics[scale=0.5]{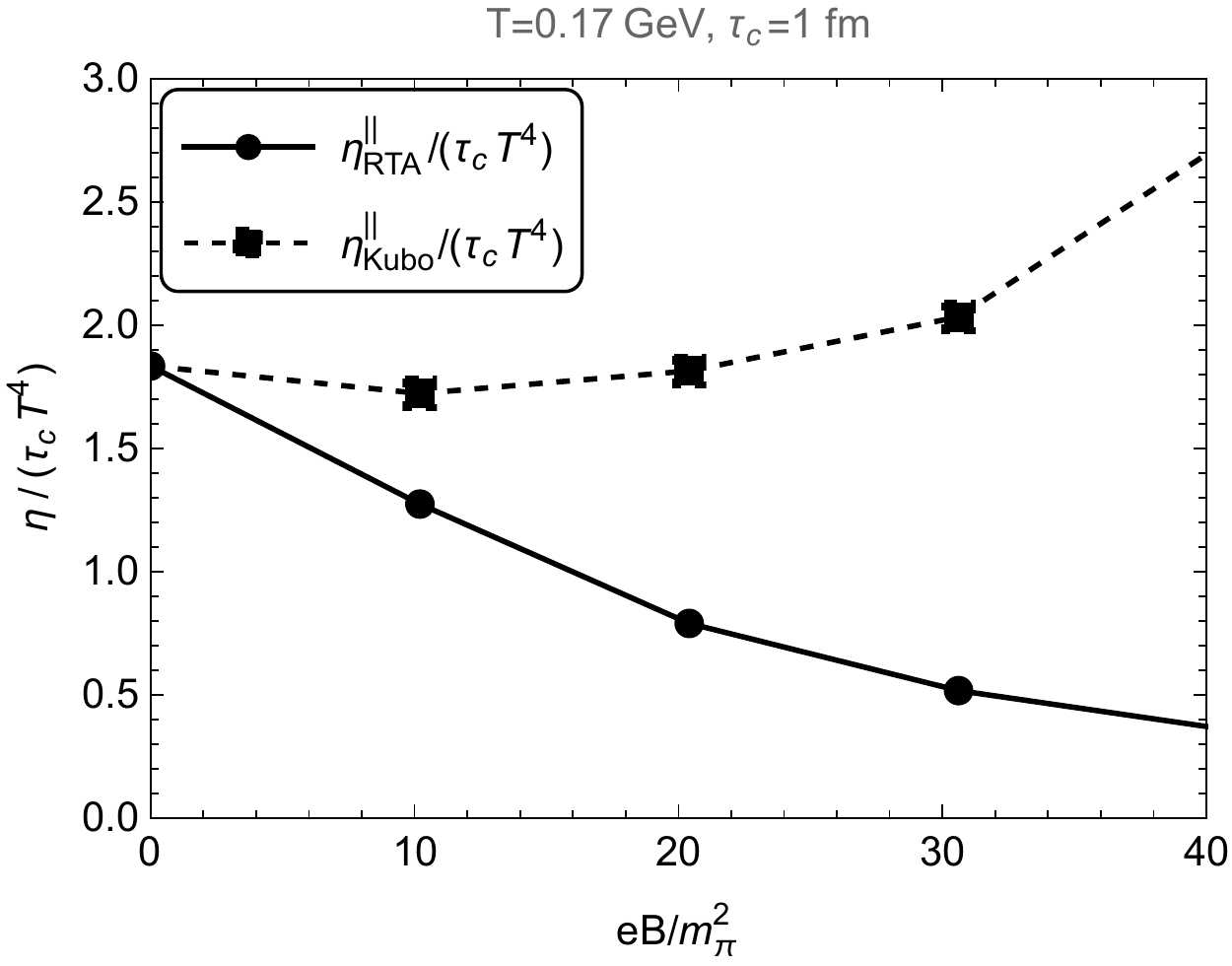}
   	\includegraphics[scale=0.5]{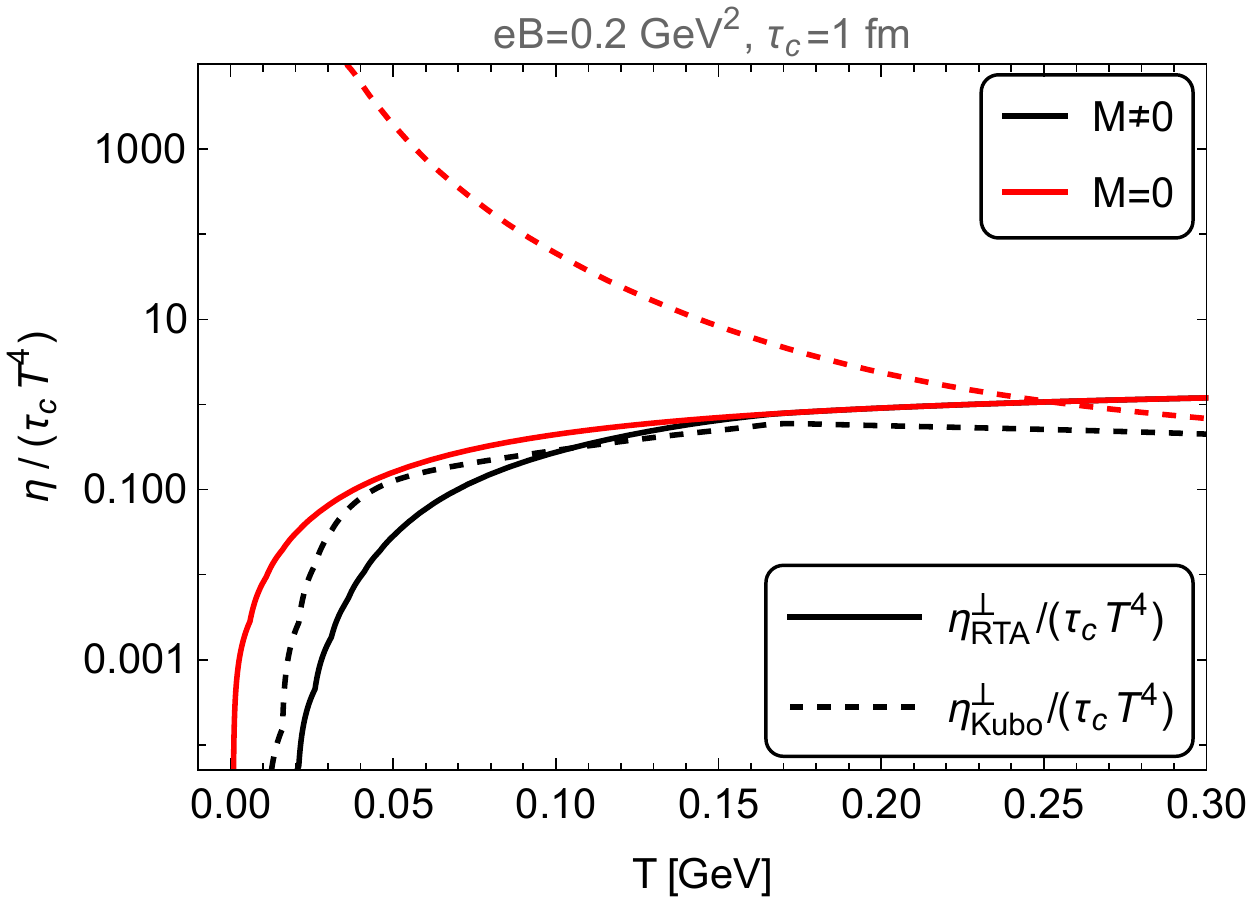} 
   	\includegraphics[scale=0.5]{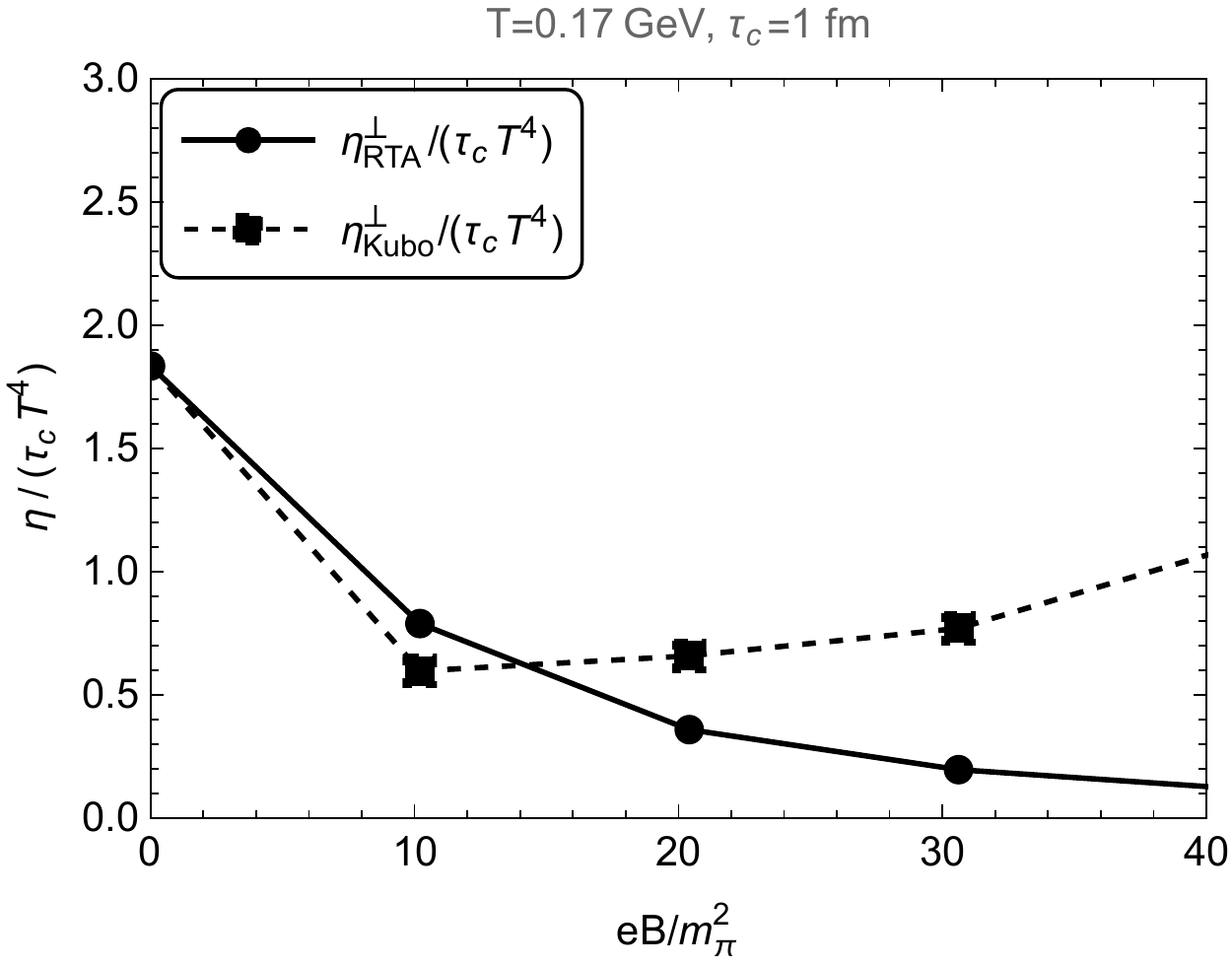}
   	\caption{Parallel and perpendicular shear viscosity of CM (dotted line), QM (solid line), QFT (dash line) curves vs temperature (left) and magnetic field (right).}
   	\label{fig:sh_TB}
\end{figure} 
Next, let us come to the other transport coefficient considered in our present study - shear viscosity, whose parallel and perpendicular components are our matter of interest. In absence of a magnetic field, both components are the same and their massless limit follows $T^4$ dependence as conductivity follows $T^2$ dependence. So we choose the dimensionless ratio $\eta/(\tau_cT^4)$, which will remain horizontal along $T$ axis for $B=0$ case. At finite $B$, RTA expressions of parallel and perpendicular components of the shear viscosity are given in Eqs~(\ref{eta_pp}), whose massless expressions can be simplified to~\cite{Dey:2019axu}
\bea 
\eta_{\parallel} &=& \frac{\eta(B=0)}{2} \sum_{f=u,d}\frac{1}{1+(\tau_c/\tau_{Bf})^2}
\nn\\
\eta_{\perp} &=& \frac{\eta(B=0)}{2} \sum_{f=u,d}\frac{1}{1+4(\tau_c/\tau_{Bf})^2}~,
\eea 
where 
\bea 
\eta(B=0)&=&\frac{g7\pi^2\tau_c}{900}T^4
\nn\\
&=&\frac{7\pi^2\tau_c}{75}T^4\approx 1.84 \tau_c T^4~,
\eea 
and same $\tau_B$'s as given in Eq.~(\ref{tB_m0}). According to above simplified expressions, normalized values of $\eta_{\parallel}/(\tau_c T^4)$, $\eta_{\perp}/(\tau_c T^4)$ of massless quark matter will increase with $T$ and decrease with $eB$. When we use constituent quark mass $M(T, B)$ from the NJL model, we will get suppressed values of RTA curves (solid line) concerning their massless curves (red solid line). The suppression represents the non-pQCD effect in the parallel and perpendicular components of shear viscosity. When we go for Kubo expressions, given in Eqs.~(\ref{eta_QFT})-(\ref{N_Sh_B}), we will get enhanced values of shear viscosity compared to their RTA values as we have noticed for electrical conductivity case. However, these enhanced Kubo values of transport coefficients are expected in the quantum domain only i.e. at high $eB$ and low $T$ domain.


At the end of this result section, we want to highlight again that the new ingredient of the present work is the Kubo estimation of NJL matter, whose RTA and QM estimations are already addressed in earlier Ref.~\cite{Ghosh:2019ubc} for electrical conductivity and Ref.~\cite{Ghosh:2018cxb} for shear viscosity. For zooming in on the Kubo contribution, the RTA and QM curves are presented here for comparison.

\section{SUMMARY}
\label{sec:sum}

Present work is aimed to highlight quantum field theoretical contribution at finite temperature and magnetic field in the transport coefficients of quark matter within the framework of NJL model. Earlier Refs.~\cite{Ghosh:2019ubc,Ghosh:2018cxb} have calculated the transport coefficients like shear viscosity and electrical conductivity of quark matter within the NJL model in the framework of relaxation time approximation (RTA). With temperature and magnetic field dependent constituent mass those estimations might be considered classical or semi-classical approaches. Present work provides their quantum field theoretical version using their Kubo expressions, obtained in Refs.~\cite{Satapathy:2021cjp,Ghosh:2020wqx}. According to the Kubo relations, one can realize the transportation from one point to another as the propagation probability at finite temperature and magnetic field of relevant field operators like electrical current $J^\mu$ and viscous stress tensor $\pi^{\mu\nu}$. On the other hand, RTA framework describe a classical picture of transport phenomena in terms of cyclotron motion of quarks due to Lorentz force. Hence in the present work we have explored this transition from classical or RTA to quantum or Kubo picture of transport coefficients in view of the NJL model. Additional temperature and magnetic field profile in Kubo estimation is the primary content of the present work, which we have subsequently compared with corresponding RTA estimations, addressed earlier in Refs.~\cite{Ghosh:2019ubc,Ghosh:2018cxb}. We also present another transition here, from massless quarks to constituent quark masses within NJL model. In low temperature and high magnetic field domain, we notice an enhancement of transport coefficients due to the transition from classical to quantum picture. On the other hand we also observe that their values got reduced during the transition from the massless case to NJL matter. Cumulative effects from both the transitions finally give us the complete field theoretical non-pQCD estimations, which remain little higher than their semi-classical values, addressed in Refs.~\cite{Ghosh:2019ubc,Ghosh:2018cxb}. These differences vanish in the classical domain, i.e.  high temperature and low magnetic field domain. { At this point we want to point out that recently some studies have been directed towards exploring the extended version of RTA~\cite{Rocha:2021zcw,Dash:2021ibx} which incorporates quantum corrections to the RTA dissipative current considered in the present study. This puts forward a good future avenue which can possibly bridge the gap between the semi-classical and the field theoretical pictures observed in the present study.}

Finally we emphasize again that the present microscopic calculation is a complete quantum field theoretical estimation of transport coefficients within the framework of NJL model at finite temperature and magnetic field. The enhanced values of the conductivity at high magnetic field domain incorporating the full quantum effect may be a good signal to show that it helps slowing down the rapid decay of the magnetic field~\cite{Tuchin:2013ie} produced in heavy ion collision experiments. But as discussed in the introduction, further studies are required to firmly conclude that.


\section*{Acknowledgments}
This work was partially supported by Conselho Nacional de Desenvolvimento Cient\'ifico 
e Tecno\-l\'o\-gico  (CNPq), Grant No. 309598/2020-6 (R.L.S.F.);  Funda\c{c}\~ao de Amparo \`a Pesquisa do Estado do Rio Grande do Sul (FAPERGS), Grants Nos. 19/2551- 0000690-0 and 19/2551-0001948-3 (R.L.S.F.); Instituto Nacional de Ci\^encia e Tecnologia - F\'isica Nuclear e Aplica\c{c}\~oes 
(INCT - FNA), Grant No. 464898/2014-5 (R.L.S.F.). Snigdha Ghosh is funded by the Department of Higher Education, Government of West Bengal, India. A.B. acknowledges the support of the postdoctoral research fellowship from the Alexander von Humboldt Foundation, Germany. Authors are highly thankful to Prof. G. Krein for his collaborative help in this work.

\end{document}